\documentclass{CVM}

\CVMsetup{
type      = {Research/Review Article},
doi       = {s41095-0xx-xxxx-x},
title     = {A Visual Modeling Method for Spatiotemporal and Multidimensional Features in Epidemiological Analysis: Applied COVID-19 Aggregated Datasets},
author    = {Yu Dong$^{1}$, Christy Jie Liang$^{1}$*\cor{}, Yi Chen$^{2}$*\cor{}, and Jie Hua$^{1}$},
runauthor = {Yu Dong, Christy Jie Liang, Yi Chen, and Jie Hua},
abstract  = {The visual modeling method enables flexible interactions with rich graphical depictions of data and supports the exploration of the complexities of epidemiological analysis. However, most epidemiology visualizations do not support the combined analysis of objective factors that might influence the transmission situation, resulting in a lack of quantitative and qualitative evidence. To address this issue, we have developed a portrait-based visual modeling method called \textit{+msRNAer}. This method considers the spatiotemporal features of virus transmission patterns and the multidimensional features of objective risk factors in communities, enabling portrait-based exploration and comparison in epidemiological analysis. We applied \textit{+msRNAer} to aggregate COVID-19-related datasets in New South Wales, Australia, which combined COVID-19 case number trends, geo-information, intervention events, and expert-supervised risk factors extracted from LGA-based censuses. We perfected the \textit{+msRNAer} workflow with collaborative views and evaluated its feasibility, effectiveness, and usefulness through one user study and three subject-driven case studies. Positive feedback from experts indicates that \textit{+msRNAer} provides a general understanding of analyzing comprehension that not only compares relationships between cases in time-varying and risk factors through portraits but also supports navigation in fundamental geographical, timeline, and other factor comparisons. By adopting interactions, experts discovered functional and practical implications for potential patterns of long-standing community factors against the vulnerability faced by the pandemic. Experts confirmed that \textit{+msRNAer} is expected to deliver visual modeling benefits with spatiotemporal and multidimensional features in other epidemiological analysis scenarios.
},
keywords  = {Visual Modeling, Epidemiological Analysis, Spatiotemporal, Multidimensional, COVID-19},
copyright = {The Author(s)},
}

\begin{document}

\maketitle

    \begin{figure}[b] \vskip -4mm
    \small\renewcommand\arraystretch{1.3}
        \begin{tabular}{p{80.5mm}} \toprule\\ \end{tabular}
        \vskip -4.5mm \noindent \setlength{\tabcolsep}{1pt}
        \begin{tabular}{p{3.5mm}p{80mm}}
    $1\quad $ & Yu Dong, Christy Jie Liang and Jie Hua, School of Computer Science, University of Technology Sydney, Sydney, NSW 2007, Australia. E-mail:  Yu.Dong-3@student.uts.edu.au, jie.liang@uts.edu.au, jie.hua@uts.edu.au.\\
    $2\quad $ & Yi Chen, Beijing Key Laboratory of Big Data Technology for Food Safety, Beijing Technology and Business University, Beijing, 100048, China. E-mail: chenyi@th.btbu.edu.cn. \\
    \end{tabular} \vspace {-3mm}
    \end{figure}

\section{Introduction}
\label{sec:intro}
A growing emphasis on the need for informatics and analytics in public health over the past 30 years has led to an increasing amount of investment cost in information systems \cite{deodhar2014interactive,carroll2014visualization}. Visual modelings in informatics raised significant roles in big data analysis tasks in public health, especially in epidemiological analysis as connections tighten up with people. However, many research challenges and gaps still need to be filled, although many new tools and algorithms have already been developed to aid experts in analyzing and visualizing the complex data used in epidemiological analysis \cite{christakis2009social}. As Lauren et al. \cite{carroll2014visualization} highlighted in a survey, complex epidemiological exploration requires novel visualization tools, and most of the existing visualizations applied in analysis tasks suffer from limited adoption. More and more standard visualizations in epidemiology evolve into visual modeling which aims for combined multidimensional features such as considerable complexity, dynamism, and uncertainty analytical tasks rather than simply spatiotemporal features. \cite{andrienko2010space,angelini2020progressive}.

As keywords in epidemiology in the past three years, the global spread of COVID-19 has overwhelmed health systems and caused widespread social and economic disruption \cite{zhang2021mapping}. Because of the virus's high transmission ability, the responses of individual countries and communities to the unfolding pandemic will be a watershed moment in human history. Although a substantial body of work has been translated into COVID-19 data visualizations globally, there hasn't been much prior work in visualization tools to conduct an analysis based on the characteristics of communities and the risk factors of COVID-19. Such intricate analysis raises significant problems related to pandemic-related uncertainty that must be addressed \cite{le2020visualising}.

To meet these challenges, we optimized collaborative approaches to problem-solving using data analytics and subject matter expertise. We conducted interviews with domain experts in health and epidemiology to gather requirements for further analysis. With the assistance of experts, a visual modeling method, namely \textit{+msRNAer}, was developed for spatiotemporal and multidimensional features in the majority of epidemiological analysis tasks. This visual modeling method adopts a portrait design inspired by viral anatomy that can be applied for visualizing each community-based location, allowing for exploration and comparison of the complex relationships among the communities' time-varying case numbers, objective risk factors that may affect pandemic, and epidemiological data patterns in fundamental geographic, timeline, and multidimensional visual designs. This creates a vivid understanding of how objective risk factors are interconnected and contributes more broadly to building resilient communities, particularly in light of the effects of pandemic transmission.

To validate the usability of \textit{+msRNAer}, we collaborated with Australian Government experts to apply COVID-19 aggregated datasets with processes of (1) extracted evidence by each local government area (LGA) from the completed 2 years in 742 days (from January 1st, 2020, to January 11th, 2022) of COVID-19 case data compressed to 106 weeks or 53 fortnights for scalability; (2) added a marked timeline with intervention events extracted by NLP; (3) connected the most recent census community profiles to each LGA; combining case data with LGA and postal area geo-locations in New South Wales (NSW), Australia. We then identified risk factors related to LGAs under expert supervision, including demographic indicators (e.g., higher-risk populations), social indicators (e.g., relationships), economic indicators (e.g., rental and mortgage affordability), infrastructure indicators (e.g., housing), and resident travel behavior (e.g., using public transportation).

We adjusted the visual designs to include multiple views and interactions, which facilitated visual exploration in COVID-19-related community portraits. This was supported by an interactive Control Panel that included event timelines, a coordinated geographic view, and a multidimensional coordinates view with a filtering function. Our aim was not only to demonstrate the effectiveness and scalability of \textit{+msRNAer}'s visual modeling but also to assist the government in investigating pre-existing community factors and discovering practical implications for potential patterns related to vulnerabilities against COVID-19. We conducted one user study and three subject-driven case studies using our COVID-19 aggregated datasets. These studies demonstrated how \textit{+msRNAer} worked with a prototype among NSW LGAs and postal areas with spatiotemporal and multidimensional features. In the user study, we not only pre-evaluated the feasibility and effectiveness of the \textit{+msRNAer} prototype but also made iterative improvements to its functions. Furthermore, the three case studies provided a selected, high-level picture of community resilience in infrastructure and explored the dimensions of resilience. We evaluated the COVID-19 exploration results and conducted interviews with domain experts to collect their feedback for future research.

The paper is structured as follows. Section~\ref{sec:related} reviews the related work. We first discuss the design requirements in Section~\ref{sec:requirements}. Based on the extracted requirements for experts, we elaborate the visual modeling with metaphor and design in Section~\ref{sec:design}. We list the data preparation and the prototype application with multiple views and interactions in Section~\ref{sec:System}. We conduct a user study in Section~\ref{sec:user} to pre-evaluate the feasibility and effectiveness of our visual modeling. We further validate the capability of our visual modeling with applied COVID-19-related case studies in section~\ref{sec:case} and interviews with domain experts in Section~\ref{sec:Discussion}. Finally, we consult the limitations and future work in section ~\ref{sec:limitation} and make conclusions in Section~\ref{sec:conclusion}.

\section{Related Work}
\label{sec:related}

In this section, we introduce related works concerning the visualization perspectives in epidemiological analysis research and accordingly, take into account how visualization aids COVID-19 spread analysis as specific examples with this background.
\subsection{Visualization in Epidemiological Analysis}
As early as 2020, a systemic review of COVID-19 epidemiology \cite{park2020systematic} proved high-spreading speed when the epidemic broke out. Since then, an increasing number of visual representations were presented to aid COVID-19 analysis. A novel study \cite{rydow2022development} on computational modeling used visualization-centric and algorithm-assisted epidemiological modeling that has proved visualization plays a critical role in epidemiological analysis in 2022. Another ongoing collaboration \cite{dykes2022visualization} between epidemiological modelers and visualization researchers summarized the concurrent challenges and solutions. They listed common visualization charts such as heat maps \cite{sanz2022modelling,pooley2022estimation} and timelines \cite{panovska2022statistical}, and further composite graphics with small multiple views which could support epidemiological modeling. Wei et al. \cite{wei2020survey} surveyed and categorized geographic visual display techniques in epidemiology research into two categories of Traditional Cartography and Geo-visualization. 

From distinct angles of data analysis, certain risk features \cite{lord2001visual,delcourt2017potential} may influence the analysis in epidemiology and make the analysis tasks more challenging. As Chui et al. \cite{chui2011visual} added human factors (age, gender, etc.) into the study of infectious diseases in the paper, further analysis of visualization in epidemiology is beneficial, and it improves the precision of algorithms like modeling and prediction \cite{roy2020prediction}. According to our classification, there are two perspectives for visualization in epidemiological analysis depending on the type of data used: spatiotemporal-based and multidimensional-based.

Pandemics are geographical in nature, and constitute spatiotemporal phenomena across large ranges of scales \cite{mocnik2020epidemics}. Improved geographic visualization plays an important role in pandemic research, that offers an environment to represent multivariate data by cartographic means, based on its geographical information effectively and attractively \cite{thony2018storytelling,goodwin2015visualizing,shapiro2017using}, and it is one of the top ten keywords of IEEE VIS \cite{isenberg2016visualization} (top conference in visualization field); also 16\% of existing related visualization works adapt maps \cite{Paul:2021:mprove}.

Multidimensional-based data analysis makes epidemiological analysis possible to connect to other factors. The Singapore epidemiology of eye diseases research is a population-based study where 8,697 adults of Malay, Indian, and Chinese ethnicity \cite{chua2017prevalence}. Steinger et al. \cite{steinger2014epidemiological} used generalized linear models to investigate the influence of key epidemiological factors on potato virus infection risk. A prototype has been deployed to demonstrate the impact of social distancing strategies during the H1N1 (swine flu) outbreak by Maciejewski et al. \cite{maciejewski2011pandemic}. Trajkova et al. analyze relevant Twitter data and discuss facilitating data interpretation via visualization to avoid the spread of misconceptions and confusion on social media \cite{trajkova2020exploring}. Also, multidimensional data visualization such as Parallel coordinates is commonly employed for visualizing multidimensional geometry \cite{inselberg1990parallel,matute2017visual,hassan2019study,zeng2021multi}. They could apply visualization research on multidimensional attributes during the pandemic, to promote the understanding of how data entries compare to each other.

\subsection{Visualization in COVID-19 Datasets}
Since the COVID-19 pandemic's initial outbreak, an astounding number of visual representations or models have been created to reflect the virus's global spread and the effects it has had on various nations and regions \cite{liu2020health,badr2020association,ulahannan2020citizen,dey2020analyzing,sanz2022modelling}. A survey study \cite{zhang2021mapping} conducted 668 COVID-19 data visualizations to map the landscape of existing visual works. Another novel research \cite{zhang2022visualization} focused on investigating complex interplay based on COVID-19 datasets between design goals, tools and technologies, data information, emerging crisis contexts, and public engagement by a qualitative interview study among dashboard creators. It could be summarized as two types of data sources in common COVID-19 research: directly-linked data such as infected cases, recovery, and mortality rates \cite{WHO:2021:WHO, Data.NSW:2021:NG}, and indirectly-linked data, which contains community information such as social impacts \cite{Sophie:2020:mccrindle}, financial impacts \cite{Alison:2021:CD,AIHW:2020:AIHW}, etc., which are not linked to the pandemic directly as objective factors. 

In common COVID-19 visualization research, basic techniques include traditional line charts, bar charts, maps, etc. Kahn's report reveals that 38\% of related works apply line/area charts, and bar charts take a 29\% share \cite{Paul:2021:mprove}. We have collected 48 existing related research works to date in Australia; 10 of them deliver a similar dashboard view, and others either offer graphs or are still ongoing works; the University of Melbourne conducts an online prototype that gives a 10-day forecast \cite{UOM:2021:UOM}. Seven projects import data from aspects such as financial, and LGA details other than only the pandemic case details. Most works apply traditional bar, stacked bar, line, map, etc. visualization methods. 

In our approach, visual modeling in COVID-19 datasets is separated into three aspects: data visualization, visual interactions, and intelligent-assistance visual analytics.

From the first aspect, raw data collected during the pandemic are applied as inputs to generate graphs. An interactive web-based dashboard \cite{dong2020interactive} to track COVID-19 in real time was first presented by CSSE at Johns Hopkins University. It is followed by an online global dashboard by the WHO \cite{WHODashboard} to show COVID-19 statuses around the globe. Hannah et al. \cite{owidcoronavirus} built 207 country profiles with aggregated cases, testings, vaccinations and etc which allow users to explore the detailed statistics on the COVID-19 pandemic. In Australia, the State and Territory governments assist the public with recognizing current statuses by visual dashboards \cite{Data.NSW:2021:NG,NTG:2021:NTG,QG:2021:QG,li2021visualizing}.

From the second aspect, visual modeling methods can combine indirectly linked data and offer deeper insights by interactively integrating multiple attributes. Dashboards with adjustable parameters are a common feature from this perspective; they combine computational analysis techniques with interactive visualizations \cite{AGDH:2021:AGDH,CA:2021:CA,USYD:2021:USYD,ANU:2021:ANU,GSA:2021:GSA,VSG:2021:VSG,bing:2021:bing,yu2021senti}, which also emphasize analytical reasoning concerning the pandemic data and other facts that may affect infection cases or get affected by the pandemic through interaction techniques. Carson et al. \cite{BigDataCOVID} presented a big data visualization and visual prototype for analyzing COVID-19 epidemiological data. Besides, more and more research has included objective factors in the COVID-19 analysis. Lan et al. \cite{lan2021geovisualization} review the usefulness of GIS-based dashboards for mapping COVID-19 prevalence and propose improved geo-visualization techniques to incorporate the temporal component in interactive animated maps to analyze pandemic transmission with COVID-19-related information. Wu et al. develop a novel Joint Classification and Segmentation (JCS) system to perform real-time and explainable chest CT diagnosis of COVID-19 infections \cite{wu2021jcs}. Muto et al. import more facts about gender, age, marriage status, poverty, and drinking/smoking habits into a matrix to address Japanese citizens' behavioral changes and preparedness against the outbreak \cite{muto2020japanese}.

From the third aspect, intelligent techniques are introduced to assist in the exploration and investigation of the COVID-19 pandemic. It takes into account more relevant yet not directly-linked complex data which includes modeling \cite{giordano2020modelling,bachtiger2020machine,regulski2021advanced,hemied2022covid}, predicting \cite{leite2020covis,chen2020visual,antweiler2021towards,xu2021episemblevis,roy2020prediction}, and other complex exploration strategies \cite{zhou2020visual,yu2022user} with AI algorithms \cite{cao2022AIcovid}. Reinert et al. developed a framework that enables effective and efficient visual exploration through interactive, human-guided analytical environments during the pandemic \cite{reinert2020visual}. Shehzad et al. purposed a decision-making environment \cite{afzal2020visual} for person-to-person contact modeling in the COVID-19 pandemic, which was based on their previous works \cite{afzal2011visual,maciejewski2011pandemic} in epidemiology. Bowe et al.'s research indicates that the pandemic plays out differently across different scales; it is related to the global supply chain, local dynamics, neighborhood mutual aid networks, and personal geographies of mitigation and care \cite{bowe2020learning}. Preim and Lawonn describe visual analytical solutions aiming to provide preventive measures. Prevention aims at advocating behavior and policy changes likely to improve human health \cite{preim2020survey}. Another research proposes a prediction of pandemic viral attack and how far it is expanding globally by Roy et al. \cite{roy2020prediction}. Guo's system discovers spatial interaction patterns, providing valuable insight into designing more effective pandemic mitigation strategies and supporting visual exploration in time‐critical situations. An approach by Christopher \cite{healey2022visual} was conveyed through visual exploration by similarity comparison and predictions. Yu et al. \cite{yu2022user} provided a user-centered visual explorer that applied COVID-19 datasets for in-process exploring and comparing spatiotemporal features in portrait-based perspectives.

\section{Design Requirements}
\label{sec:requirements}

In 2022, a survey conducted by Jason et al. \cite{dykes2022visualization} summarized the challenges, solutions, reflections, and recommendations of visualization for epidemiological modeling. The authors categorized the supporting visual modeling in epidemiological analysis into three stages based on different time scales:

The initial stage involves the quick application of candidate templates with visualization tools to establish problems. Data is transferred to preset combinations of views for simple comparison.

In the short-mid term stage, ongoing research provides a redesign of the visualization prototype for iteratively redefining the problem, exploring potential patterns, and providing users insight into complex tasks.

In the long-term stage, a more stable visual system is developed for widespread application to common usage scenarios, as demonstrated through multiple cases in epidemiological analysis.

This summary underscores the motivation for our proposed visual modeling method in epidemiology to fulfill multiple objectives. The visual modeling method for epidemiological analysis should not only enable quick responses to basic information trends of a pandemic, such as the infection cases in time-varying trends but also facilitate the discovery of knowledge concerning combined analysis tasks, such as identifying the factors that affect infection cases. Finally, the visual modeling method must be applied to a completed visual system and validated using real-world cases.

In consultation with experts from the epidemiology and health domains, we outlined the following user requirements: Experts required information to assess the profile of communities in terms of their resilience to virus attacks. Combined with the inspection of the number of infected cases, a few key factors might affect and help understand the impacts of different pandemic phases occurring concurrently. They must assess the impact of the government interventions and measure the resulting pandemic situations in order to investigate both the community profile and the infection cases concurrently. They needed to investigate the link between community factors and the number of infection cases caused by outbreaks, intervention events, and responses. Usability studies conducted with early existing prototypes identified a variety of requirements. The desired features, which we distilled into three progressive categories of peer-to-peer design requirements for our approach from aspects of visual design, visual analytics, and modeling prototype application during several design iterations with subject matter expertise.

\textbf{R1. Provide comparative visual portrait design of the numerical distribution of consecutively transmitted cases for each community:} To perform effective epidemiological analysis, it is essential to describe consecutive infection cases in terms of timelines and locations. This involves investigating the geographic and temporal trends of the epidemic's spread, along with qualitative analyses utilizing spatiotemporal features. By providing comparative visualization portraits of these aspects for each community, we could better understand the patterns and trends of transmission.

\textbf{R2. Offer visual exploration for analyzing transmission patterns with spatiotemporal and multidimensional features among each community:} Epidemic outbreaks are often related to various objective factors in specific locations, such as socioeconomic or cultural factors. Therefore, it is crucial to support multidimensional feature exploration, including humanities and finance, among other fields, in addition to visualizing the spreading situation with spatiotemporal features. By combining both perspectives in interactive portraits of location-based risk factors and infection cases over time, we can provide a more comprehensive understanding of transmission patterns.

\textbf{R3. Verify the effectiveness of visual modeling through a prototype system using actual epidemiological cases:} To demonstrate the effectiveness of our visual modeling approach, the prototype system should offer multiple visual views with robust interactive functions, such as collaborative filtering and comparison. By applying this system to real user studies and case studies in epidemiological tasks, we can show how our approach can help researchers gain insights and make informed decisions in the field.

\section{Visual Modeling}
\label{sec:design}
To address the design requirements, this section presents our design rationale through visualization, design considerations, and guidelines for the components.
\subsection{Design Metaphor}
\begin{figure}[h!t]
\centering
\includegraphics[width=3.5in]{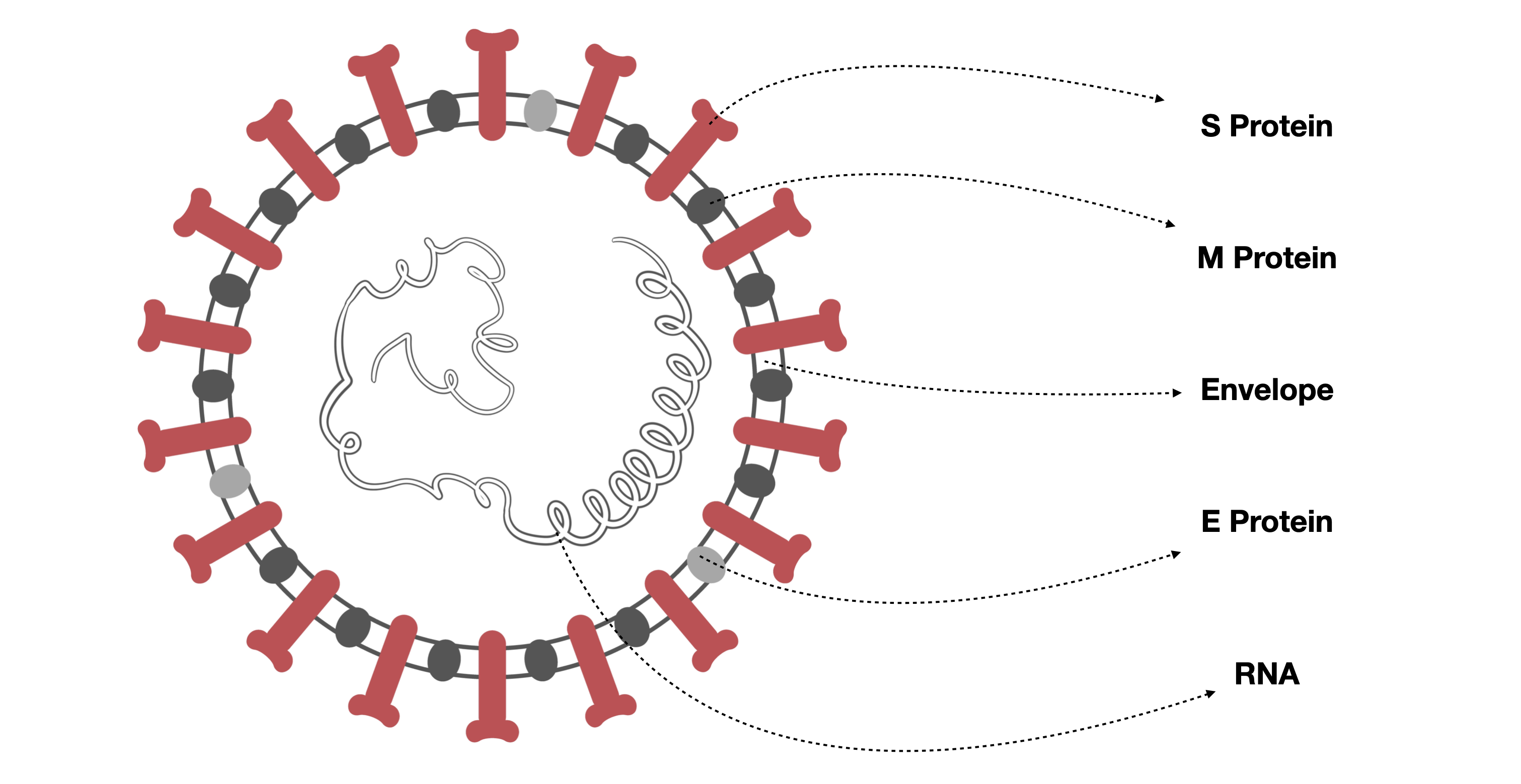}
\caption{The cross-sectional simulation of SARS-related coronavirus with its main components.}
\label{fig:Virus}
\end{figure}

Inspired by viral anatomy, our visual design primarily adopts the 2D genome structure of SARS-related coronavirus particles. Coronaviruses, named for their "crown-like" shape observed in the electron microscope, have particles packed with shells. This illustration became familiar to the public, as shown in Figure ~\ref{fig:Virus}, and is widely introduced by the news and media. The coronavirus particles are organized with \textit{+sRNA} (positive single-stranded RNA) polymers packed inside, further surrounded by outer inserted proteins. These outer proteins derive from the cells in which the virus is last assembled but are modified to contain specific viral proteins, including the Spike (S), Membrane (M), and Envelope (E) Proteins. The S Protein allows viruses to enter and infect other cells. After the virus enters the host cell, the genome was transcribed, and replication takes place involving coordinated processes of RNA synthesis. Positive Multiple Strands of RNA Encoder, abbreviated as "\textit{+msRNAer}", is the name of our proposed visual modeling method. Inspired by this simulation and combined with our previous research \cite{Pansytree2020,yu2022user}, we map different parts of the virus into multiple meanings and design \textit{+msRNAer} to apply the aggregated datasets of its inner and outer parts.

\subsection{Visual Design}
\begin{figure}[h!t]
\centering
\includegraphics[width=3.3in]{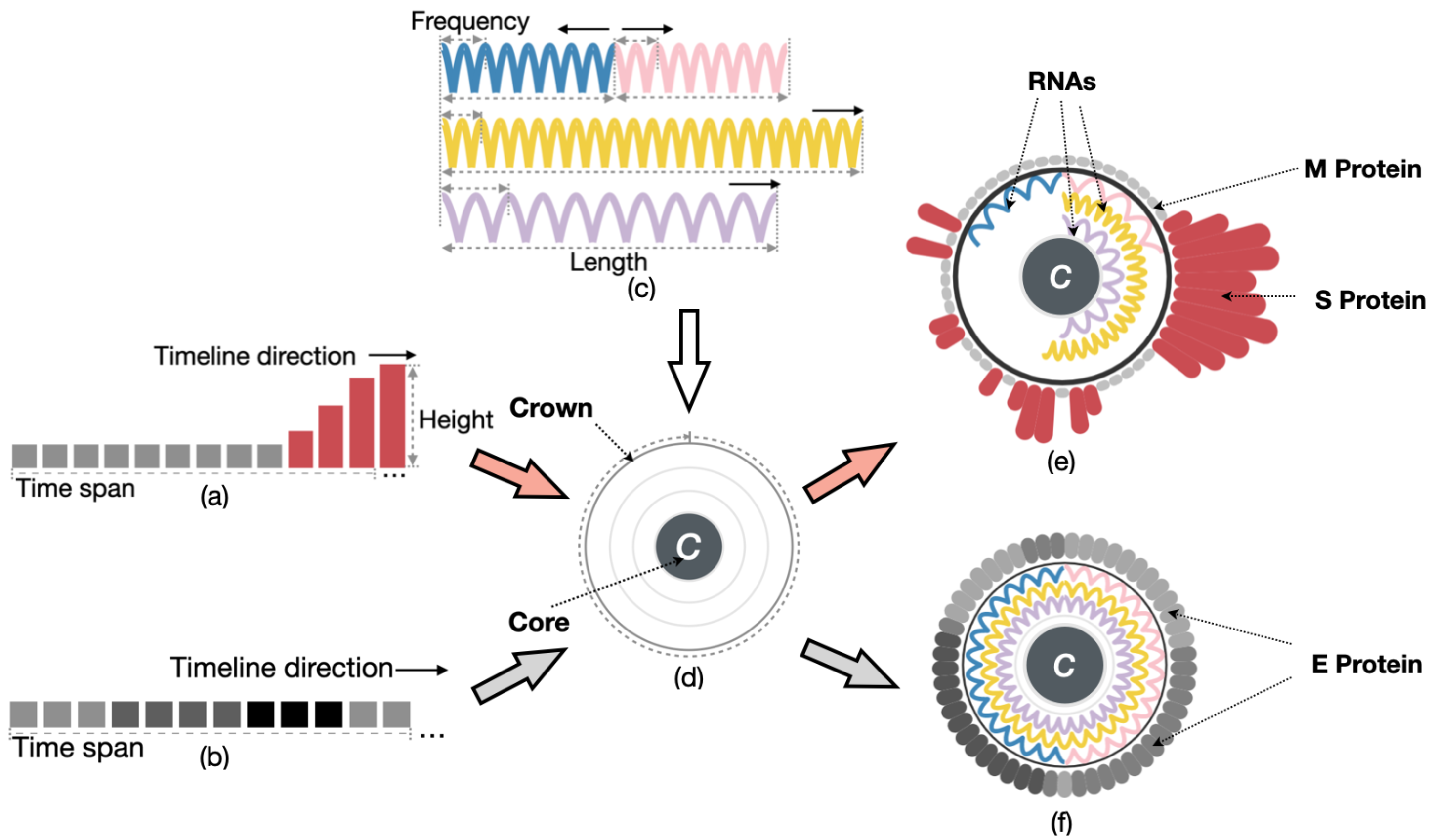}
\caption{Two types of visual portrait and their implementation processes: (e) Portrait with S and M Proteins, and (f) Portrait with E Proteins. Both Portrait (e) and (f) include 4 strands of RNAs. The red arrow path represents the process of (a) combining bars on the timeline and encoding them as S and M Proteins, while the light gray arrow path represents the portrait with (b) E Protein in grayscale, depicting different phases of the timeline.}
\label{fig:Bar}
\end{figure}

We depict a novel visualization that leverages the biological components of the coronavirus as metaphors to represent and compare communities' epidemiological characteristics. Specifically, we define the particle symbol for each community, which we call a "Portrait" and designed the outer Proteins and inside RNAs to encode information related to cases (such as actual cases, cases per 10k/100k population) and transmission trends and each community's unique risk factors respectively. Figure ~\ref{fig:Bar} is the detailed design for all components. Notably, we have varied the RNA from positive single strand to positive multiple strands to capture multiple key risk variables related to community characteristics. This design choice provides multidimensional insight into the epidemiological patterns of each community and aims to fulfill R1.

As domain experts suggested, the number of infection cases caused by the pandemic is related to multiple potential factors among communities, for example, the population and their percentage in higher-risk groups, including the aged, lower-income, and lone-person groups. As a result, we could emphasize community portraits based on the characteristics of the vulnerable population in high-risk areas. Hence, we assemble all visual elements as the portrait for each community, as shown in Figure ~\ref{fig:Bar}. This single aggregated crown-like portrait describes both the community's existing characteristics and also how the community reflects viral infection case numbers information. The visual portrait consists of three components: the crown, outer designs with Proteins, and inner designs with RNAs. The circle of a crown represents one unit of the whole timeline. On the outside, we use multi-segments as time spans, with S Protein representing case numbers information and M Protein representing zero cases, and multiple RNAs representing the highlights of selected factors in higher risk groups inside the crown. The E Protein, unlike the S and M Proteins embedded on the crown, can represent significant intervention events rather than case numbers. 

\textbf{Crown Design}: We define the central core with radius $R_c$ of the Crown to label the community $C$, as shown in Figure ~\ref{fig:Bar}(d). The circular loop surrounding the Crown easily depicts the entire timeline of any transmission taking place. The circle is evenly divided into continuous segments matching time spans for S or M Proteins clockwise from the top point to the looped end. Or, on the other hand, we use E Proteins inserted to denote the categorized intervention events within the timeline colored in grayscale.

\textbf{Outer Design}: The outer-growing S Proteins are densely packed together. For each community over each time span, we encode distinct bars with rounded corners, with their heights representing the case number shown in Figure ~\ref{fig:Bar}(a). We use the smallest-sized M Proteins to indicate zero cases in a certain time span. The S or M Protein height of time span $x$ is calculated as follows:

\begin{equation}
H_{timespan_{x}}=\left\{\begin{array}{ll}
h+{R^{\prime}_{c}} * (a+\ln f(x)) * b & , f(x) \in N^{*} \\
h & , f(x)=0
\end{array}\right.
\end{equation}

A base height $h$ appears when there are 0 cases in one time span, and a growth height is proportional to the number of non-zero infection cases in a time span $x$, in which $f(x)$ is the function of infection cases corresponding to this time span. Both $a$ and $b$ are customized parameters, and $R^{\prime}_{c}$ refers to the initial radius of the Crown.

\textbf{Inner Design}: 
Although viruses normally only carry one strand of RNA, it is not ideal to employ only one strand to represent several important elements. We suggest splitting the RNA strand into numerous strands to show these parameters in order to conserve space and reduce unnecessary requirements. Thus, we propose to distribute the four RNA strands across three channels rather than joining them head-to-tail. Each element's maximum value is set to occupy half of a channel when multiple data elements of the same type must be represented, such as visualizing the numbers of males and females (exclusive of transgender people). These two can then be combined and allocated to a single channel. Other whole channels may be assigned to data of a single type. Figure \ref{fig:Bar}(c) demonstrates the potential of our visual design to share one channel with two RNAs. We implement three channels of RNA that grow from the middle point of the circle. Combining two related factors in one shared channel accommodates all four RNAs placed in three-oriented channels within the Crown.

To fulfill the visual design, we employ a visual metaphor of spiral genomes with four cosine wave-shaped RNAs with the same amplitude in three channels. As shown in Figure ~\ref{fig:Bar}, the length of RNA is encoded by the exact values (e) of its categories.

To denote the length of RNA $L_{ij}$ in the arc, we define: 
\begin{equation}
L_{ij} = \left[{R_c} + (m-0.5) * \frac{{R^{\prime}_{c}} - {R_c}}{3}\right] * ({\frac{N_{ij}}{max\{N_i\}}*\frac{2\pi}{n} + \frac{\gamma}{n}})
\end{equation}   

I.e., the arc angle $\theta_{ij}$ corresponding to the arc length $L_{ij}$ is: 
\begin{equation}
\theta_{ij} = {(\frac{N_{ij}}{max\{N_i\}}*\frac{2\pi}{n} + \frac{\gamma}{n})} * {\frac{2\pi}{2\pi + \gamma}}
\end{equation}

Equation (3) employs a rescaling that ensures the maximum will not exceed the current RNAs' located channel lengths. $N_{i}$ represents the set of data from the independent variable RNA category $i$, then ${max\{N_i\}}$ denotes the maximum value in $N_{i}$. $N_{ij}$ refers to the value of data from an independent variable community $j$ in an independent variable RNA category $i$. Parameter $m \in \{1,2,3\}$ allocates the exact location in the channel of the current RNA category $i$. Parameter $\gamma$ is the minimum arc angle in RNAs, maintaining RNA even if the value of $N_{ij}$ is tiny. Parameter $n \in \{1,2\}$ indicates whether the maximum length of RNAs occupies a full channel or a half channel. We define $\Theta_{ij}$ as the maximum arc angle with $\Theta_{ij} = \frac{2\pi}{n}$ when $N_{ij} = \max\{N_i\}$. Therefore, any $\theta_{ij} \in [0, \Theta_{ij}]$.

Further define the cosine wave function $F_{ij}$ for drawing each RNA as: 
\begin{equation}
F_{ij} = \frac{{R^{\prime}_{c}} - {R_c}}{3} * \left[|\cos(\theta_{ij} * \frac{N_{ij}}{N_j} )| + (m-1)\right]
\end{equation} 

Equation (4) is simplified by the designed function $F_{ij} = \frac{{R^{\prime}_{c}} - {R_c}}{3} * |\cos(\theta_{ij} * \frac{N_{ij}}{N_j} )| + (m-1) * \frac{{R^{\prime}_{c}} - {R_c}}{3}$. This equation is assembled from two parts: one draws the cosine shape with an absolute value function, while the other is used for radial translation. The parameter in equation (4) applies $\frac{{R^{\prime}_{c}} - {R_c}}{3}$ as amplitude after absolute value calculated, ${N_j}$ as the sum value of data from community $j$, then the frequency in cosine waves $\frac{N_{ij}}{N_j}$ represents the ratio of the current value from community $j$ in RNA category $i$ divided by the sum value of community $j$. The above RNA design allows for the comparison of different communities within the same RNA category from two distinct perspectives. The first aspect involves comparing current categorical values among multiple communities based on RNA length, while the second aspect relates to comparing the ratio of current values with the total value in the same community, as determined by RNA frequencies.

As an illustration, the RNA for selected factor $income$ from community $Sydney$ should be mapped when allocated in the second channel: $F_{income, Sydney} = \frac{{R^{\prime}_{c}} - {R_c}}{3} * |\cos(\theta_{income, Sydney} * \frac{N_{income, Sydney}}{N_{Sydney}})| + \frac{{R^{\prime}_{c}} - {R_c}}{3}$.

\textbf{Filter Trigger Design}: All the portraits are needed to motivate a filter trigger. We reset the Crown and created a sample portrait with E Proteins for filtering purposes, which can be used interactively to attach events to the timeline. On the outer circle, the grays on the circle indicate different events by timeline. In the inner Crown, three full-circled values of RNAs, as shown in Figure ~\ref{fig:Bar}(f) are the indicators for selected factor groups. During exploration, we intend to interact with the visualization by interacting with all these elements in Control Panel.

\textbf{Design for All Colors}: Two sets of color scales for Control Panel and Portrait View have been used in our visual design. To raise awareness of the threat, we encode bright red for the S Protein design, encoding the infection case numbers, and light grey for the M Protein, indicating there were no cases this week. Inside the Crown of the portrait, inner color scales are used to representatively paint community portraits on RNAs, which include azure blue, mint pink, gold yellow, and pale purple. In Control Panel, a pre-defined grayscale is designed for E Protein to differentiate different types of intervention events along the timeline, initially with normal gray, silver gray, and dark gray. The darker shade of gray indicates a higher level of restriction for the events. For interaction, charcoal gray is used in both Control Panel and the portrait design for selection interaction.

\section{Data Preparation and Prototype Application}
\label{sec:System}

We further developed a web-based visual prototype based on \textit{+msRNAer}, which aimed to assist in investigating the pre-existing community factors and discovering practical implications for potential patterns of established community characters against the vulnerability faced by epidemiological analysis.

Although \textit{+msRNAer} should be suitable for most epidemiological analyses with community factors, we introduced the aggregated COVID-19 datasets in NSW as hot issues from epidemiological analysis to better present the design ideas in the previous section. We then applied \textit{+msRNAer} to the prototype implementation by each view and introduced the interaction designs in this application.

\begin{figure*}[!t]
\centering
\includegraphics[width=6.8in]{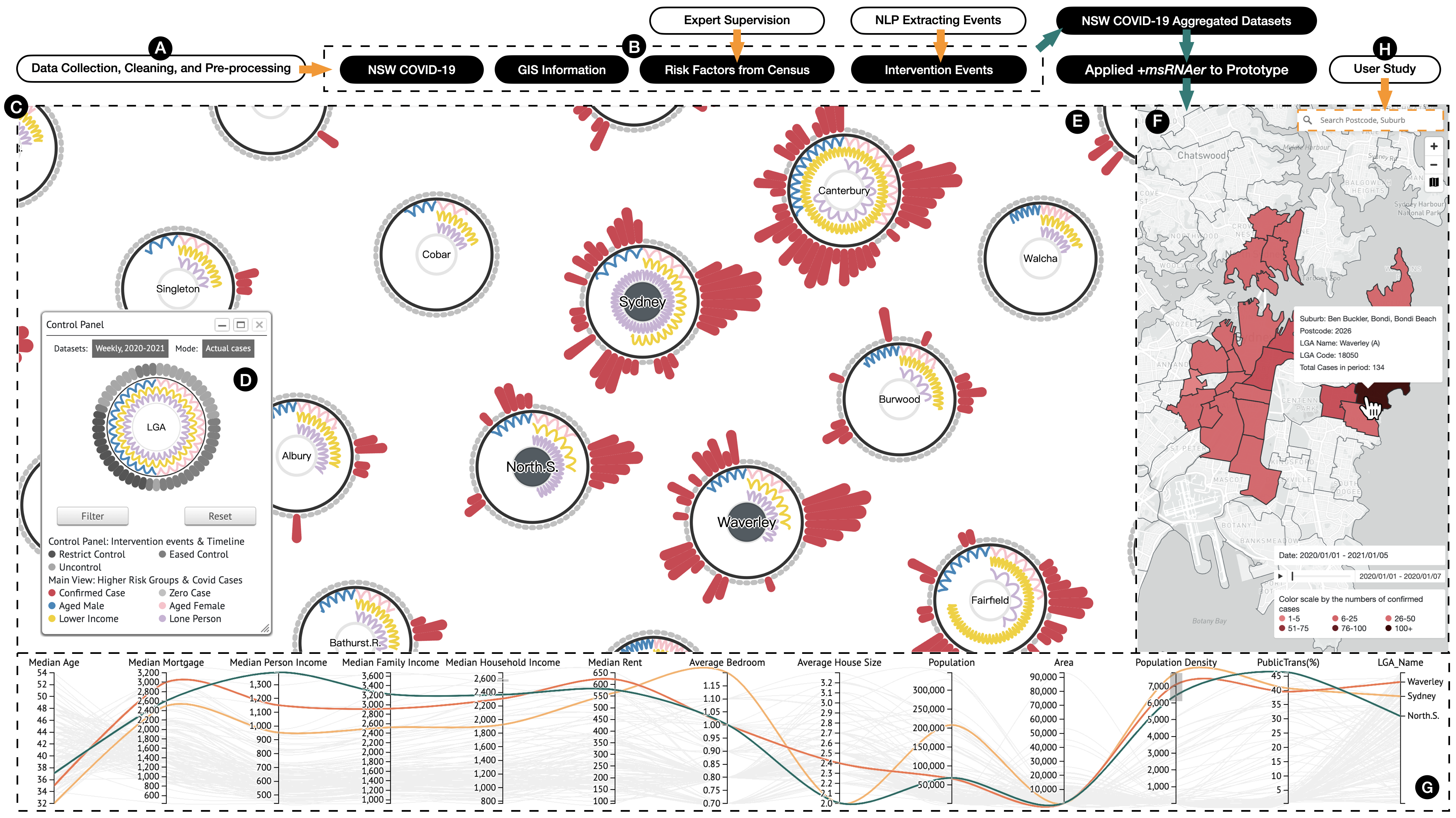} 
\caption{The entire workflow of the visual modeling method \textit{+msRNAer} applied to NSW COVID-19 aggregated datasets, from raw data collection to prototype system which embodies one completed pathway. It consists of two types of arrows, where the tangerine color connects the human-participated actions and the cyan color means the data flow direction. The pathway starts from (A) data collection and processing, (B) dealing with datasets summarized COVID-19 cases, GIS information, LGA-based censuses under experts' supervision, and intervention events extracted by NLP, and assembling the aggregated datasets for application in \textit{+msRNAer}; (C) the prototype interface with (E) Portrait View depicts four high-risk factors and COVID-19 cases corresponding to the selected timeline in (D) Control Panel and interacts with (F) and (G); (F) the GIS View with highlighted polygons of LGAs and postal areas shows their COVID-19 distribution; (G) the MDC View is a high-level overview of other risk factors from the census indicators. This \textit{+msRNAer} prototype is finally completed by a conducted user study by adding a search function (H) in GIS View.}
\label{fig:workflow}
\end{figure*}

\subsection{Data Sources}
We intended to investigate whether there had any pattern or potential relationship between NSW COVID-19 cases and demographics, geo-information, infrastructure information, or other factors in the census fields, as clarified by the requirement of analyzing COVID-19 situations. Multiple datasets were prepared to aggregate for demonstrating the effectiveness of our approach, which includes COVID-19 case data with event timelines and selected columns of NSW census data.

The COVID-19 case data in NSW was collected by the government, and the pandemic data program led by the Data Analytics Centre (DAC) provides digital information to improve the coordination of the government’s COVID-19 response. The datasets provided applicable information about infection cases based on the location of usual residence since the first infection case; they excluded 189 cases in crew members who tested positive while onboard a ship docked in NSW at the time of diagnosis; and case aspects include confirmed, tested, recovered, and death by their notification date, location, age-range, and likely source of infection. Some of them were no longer released due to privacy. Plus, Geographic Information System (GIS) map data and event information were respectively extracted from media releases \cite{NSWHealthMedia} and NSW Property Web Service \cite{NSWGeo} authorized by NSW government websites.

We decided to identify the high-risk factors that may aid in COVID-19 infection from the census data, which involves millions of people and households and is conducted by the Australian Bureau of Statistics (ABS) every four years. The data provided a rich snapshot of the nation and informs the government, communities, and businesses; it contained essential concepts such as populations, rents, mortgages, incomes, religion, languages, housing, and more. Besides, based on the definition from ABS, a lone person is classified as the only person aged 15 or over who lives in a private dwelling. Also, people in NSW earning more than 50\% but less than 80\% of the NSW or Sydney median income are described as earning a lower-income \cite{NG:2019:NG}. The next wave of census data has been released in stages since June 2022, but detailed information will not be released until mid-2023. Thus, the latest census data from 2016 was applied in this paper.

\subsection{Variables Consideration}
The COVID-19 case dataset delivers every case recorded by NSW Health and contains multiple attributes concerning cases by notification date and postcode, local health district, LGA, and likely source of infection. Considering the risk of leaking information that could directly identify individuals, only personal age, gender, and the location of their usual residence about infection cases are included in this dataset, which is assessed to measure the risk of identifying an individual and to measure the information gained if it is known that an individual is in the dataset. LGA is an official spatial unit that contains multiple postal areas that represent the whole geographical area, and there are 128 LGAs in NSW in Australia (Bayside Council was formed on September 9th, 2016 after the 2016 census, and relevant LGA data is merged from the City of Botany Bay and the Rockdale City Councils). Moreover, as data journalism may affect the COVID-19 pandemic \cite{desai2021data}, relevant news articles, alerts, and ministerial media releases issued by the NSW government about COVID-19 are attached as events to combine with the notified date of infection cases.

The timeline selection was intercepted during the COVID-19 pandemic, from January 2020 to January 2022. We used NLP \cite{mikolov2013efficient} (Natural Language Processing) to extract textual information from media data and considered marking the timeline related to keywords in interventions and social restrictions as three types of phases: uncontrolled, eased (e.g., keeping social distance, masks required), and restrict controlled (e.g., curfew, bubble restriction, lockdown); and the marked timeline was attached to the COVID-19 cases dataset for two data periods: The long period used 53 fortnights of COVID-19 case data as a biweekly time span grouped by 106 weeks of data from January 1st, 2020, to January 11th, 2022; the short period was listed 53 fortnights of case data as a weekly time span from January 1st, 2020, to January 5th, 2021. Further imported JSON files combined a single LGA layer for the long period and an LGA layer with postal areas for the short period of COVID-19 case data.

We further consulted domain experts and selected the four top key factors from a suite of factors by their supervision that may cause infection in their communities (LGAs and postal areas) as directed: aged males and females (70+), lower income groups, and living alone groups. Experts extended four additional categories of indicators from censuses that may affect the COVID-19 pandemic: LGA-based demographic indicators, social indicators, economic indicators, infrastructure indicators, and resident travel behavior. Each category contained over 30 indicators and descriptions. Some indicators contained complex hybrid patterns that may influence transmission. For instance, apart from living alone, the indicator of household is defined as different from a family, which refers to at least one person over the age of 15 who lives in the same private dwelling. In this situation, it is necessary to deliberate details on the comprehension of demographic, social, and economic indicators. We extracted the data on median age, population, and area size from LGA demographic information and further calculated population density; median rent expense, median mortgage, median personal income, median family income, and median household income from social and economic indicators; the average bedroom per person and the average bedroom size per household as dwelling factors from infrastructure indicator; and the public transportation rate for traveling from resident travel behavior.

Finally, we finalized 96460 rows of COVID-19 case data up until January 11th, 2022, and aggregated both datasets with intervention events and spatiotemporal features as well as simplified objective factors from the NSW census data tables weighing 449 MB.

\subsection{Prototype Implementation}
The prototype was implemented to meet requirements that offer extra visual explorations of interactive community portraits. For common epidemiological analysis tasks, we assembled to propose \textit{+msRNAer} prototype with collaborative visualization views, as shown in Figure~\ref{fig:workflow}, including Control Panel (D), Portrait View (E), Geographic View (F), and Multidimensional Coordinates View (G), implemented based on D3.js \cite{bostock2011d3} and Mapbox.js \cite{Mapbox2022} as base map.

For better illustration, we applied our data preparation to the NSW aggregated COVID-19 dataset. The entire workflow of the applied \textit{+msRNAer} prototype is demonstrated in Figure ~\ref{fig:workflow}. In this subsection, \textit{+msRNAer} prototype is introduced as being integrated into the pipeline workflow with summarized human-participated actions and data flow directions by steps, which include collecting and dealing with raw datasets, processing specific data with NLP, and extracting information under expert supervision to the aggregated COVID-19 datasets, and applying \textit{+msRNAer} to the prototype system with several functions and interactions finalized.

\textbf{Control Panel and Portrait View.} The Portrait View is built containing a Control Panel based on our visual modeling method with a force-directed layout to avoid overlapping and improve readability. First, we can select the applied datasets and inspection of modes - either actual cases or cases per 10k population - in the Control Panel. Then, we can hover or click on each time span on the sample portrait to interactively observe the corresponding filter result on each portrait and inspect all color tooltips among each view.

We can further inspect, explore, and compare infection cases and community key factors for each in the Portrait View, with aggregated case numbers by time span on the Crown and overviews of selected risk factors in the inner. We fix Portrait View as the main view and we set each portrait with the first channel with azure blue for older males and mint pink for older females, the second with gold yellow for lower-income groups, and the third inner channel with pale purple for lone people groups.

\textbf{Geographic (GIS) View.} We added extra visual polygonal layers to the Mapbox-drawn landscape in order to divide different LGAs and postal areas in NSW. When an LGA or postal area is selected, it will be highlighted with a charcoal gray polygonal boundary and will display detailed information about geographical information and infection cases. Using the GIS View, we can explore geographical information by zooming in on postal areas within the LGA and zooming out for an overview. The GIS View also provides a playable timeline window for enabling the inspections of the spreading situations in selected time periods. The visual design of GIS View utilizes a red gradient scale. The darker the red, the more infection cases are represented in each LGA.

\textbf{Multidimensional Coordinates (MDC) View.} We utilized parallel coordinates-based visualization for multidimensional variables. The MDC View aims to show a high-level overview of other factors related to community resilience and allows users to explore the dimensions of resilience. Each LGA polyline is distinguished by coordinates, different colors, and different styles. Polyline uses different dash styles or colors according to the positive proportional to the infection cases.

\subsection{Visual Interactions}
We can interact with multiple views, including Portrait, GIS, and MDC Views, triggered by Control Panel. We offered flexible interactions within and among multiple views collaboratively on this visual prototype system for each visual exploration from multiple perspectives, summarized as follows.

By combining the Control Panel and the Portrait View, we have implemented several interactions that enable efficient navigation through the LGA portraits and visual cues within the core for further exploration and comparison.

\textbf{Filtering by Time Spans.}{}
The interactive Control Panel serves as a starter to filter LGAs in all visual views and by time spans based on intervention events and to address all color legends in the Portrait View. Users can independently inspect the COVID-19 cases on S Protein or M Protein among each LGA portrait based on the gray-scale mapped on E Protein, a time period of intervention events, or a time period before and after the restriction events.

\textbf{Zoom and Pan.} An interactive portrait based on SVG supports zooming and panning for exploring overviews or detailed visual elements.

\textbf{Drag and Reposition.} The force-directed layout ensures that each portrait does not overlap. Users can drag and lock any portrait to a new location to allocate any selected LGA for comparison. In addition, we enhance interactions for further functionality by left-clicking to mark the cores and canceling their current location by double-clicking.

\textbf{Highlight Visual Cues and Switch Contexts.} When hovering the mouse over each type of visual cue in the Portrait View, corresponding visual cues and associated data information tooltips will be highlighted for further inspection.

\begin{figure}[h!t]
\centering
\includegraphics[width=3.3in]{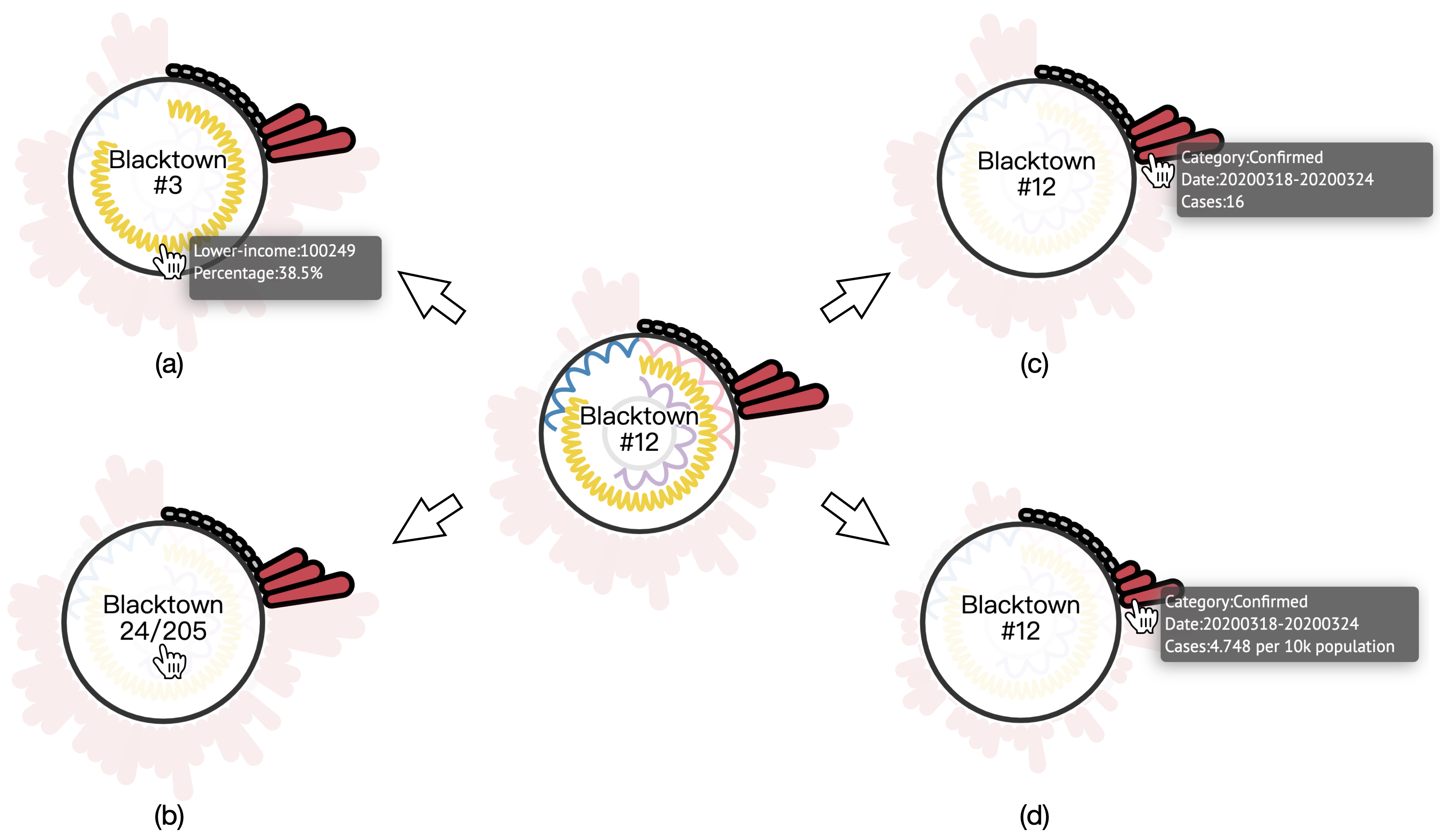}
\caption{Four highlighting strategies are displayed when hovering over: (a) the ranking and tooltip of the selected RNA among all LGA portraits, (b) the sum of case numbers in the filtered phase with the total cases in the entire period, (c) the ranking and tooltip of case numbers summed by hovering time span, and (d) the context is switched to the case numbers per 10k population by hovering time span.} 
\label{fig:Interaction1}
\end{figure}

\textbf{Conjunction with filtering interaction}, as shown in Figure ~\ref{fig:Interaction1}, it supports dynamic highlighting of ranking or case numbers in the core based on the cases or prevalence rate with the average cases per 10k population in the previous filtered timeline added with the hovering S Protein's height. Alternatively, users can hover only on RNA to highlight its ranking among all LGAs.  

The GIS and MDC Views collaborate with the Portrait View by highlighting the selected both LGAs and postal areas. Some other interactions are also supported.

\textbf{Playable Timeline Window.} The GIS View provides an interactive timeline window that allows users to investigate the transmission situation over different time spans. It includes several functions such as auto-play and pause, enabling users to customize their explorations.

\textbf{Boundaries Highlighted.} The GIS View supports clicking to highlight boundaries and reflect LGA portraits in the Portrait View. It also supports connecting with the cores in the LGA selection.

\textbf{Heatmap Highlighted.} The GIS View offers a heatmap layer in LGA or postal areas by timeline filtering in Control Panel. The colors reflect the number of COVID-19 cases in the selected time period.

\textbf{Brush on MDC View.} In MDC View, the brush function is used to filter the portrait factors in multiple dimensions and reflect them to other views. 

\section{User Study}
\label{sec:user}
To ascertain the viability and effectiveness of \textit{+msRNAer} prototype, we conducted a user study before applied to case studies. In this section, we presented the details of the study setup and analyzed the obtained results.

\subsection{Participants and Apparatus}
In our user study, we endeavored to recruit a diverse group of participants with varied backgrounds and levels of research experience in the field of computer-related disciplines. Ultimately, we successfully recruited 16 volunteers from our campus, comprising an equal representation of 8 male and 8 female individuals. Notably, the age range of our participants was wide, ranging from 19 to 30 years, with a mean age of 24.06 years. Additionally, we verified that 5 of our participants were enrolled college students without any prior research experience, while the remaining 11 were postgraduate candidates with research experience ranging from one year to eight years, resulting in a mean of 1.88 years of research experience. Despite their shared interest in visualization, all participants reported being unfamiliar with visualization methodologies.

All user studies were planned to be presented in the campus study pods. Our \textit{+msRNAer} prototype was supported by an Apple MacBook Pro (15in, 2018) equipped with 16GB of memory, i7 processors, and a Radeon Pro 555X Graphics Card which provided participants for visualizing and interacting with \textit{+msRNAer} prototype clearly and effectively on a 60-inch LED external monitor with 1920 x 1080 resolution.

\subsection{Tasks}
The tasks included in our user study incorporated quantitative and qualitative analysis. The quantitative analysis provided objective numerical data, while the qualitative analysis allowed us to gain subjective insights into participants' interactions that aimed to verify the feasibility and effectiveness of \textit{+msRNAer} prototype.

\begin{figure*}[!t]
\centering
\includegraphics[width=6.8in]{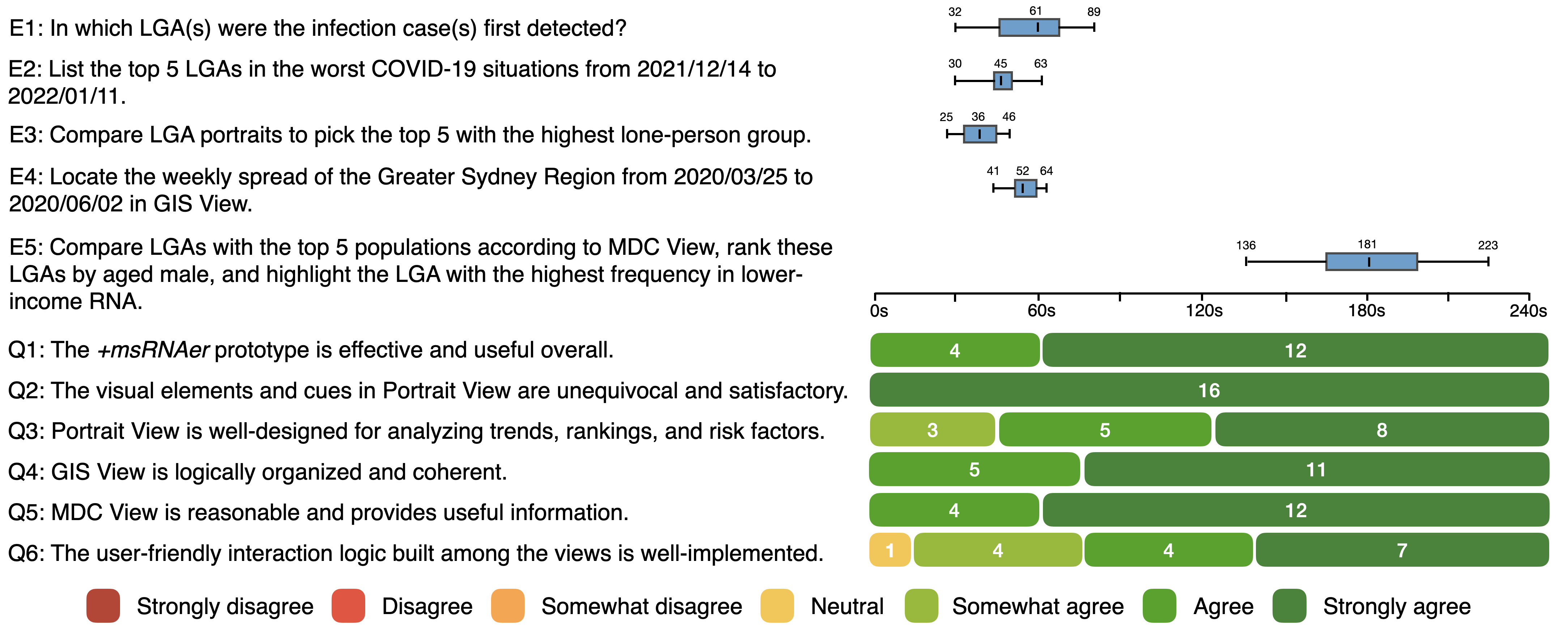} 
\caption{The quantitative analysis results of the exploration task list (E1–E5) and the questionnaire (Q1–Q6), where exploration tasks record the completed time in a box plot and the questionnaire counts participants' choices in stacked bars.}
\label{fig:userstudy}
\end{figure*}

To enhance clarity in quantitative analysis, we established specific exploration tasks for each view as well as collaborative tasks among the views to test the exploration capabilities of \textit{+msRNAer} prototype. By recording the completion time of each task, we intended to conduct analyses to determine whether \textit{+msRNAer} prototype improved the exploration capabilities for portraits' COVID-19 case trends and risk factors and whether they had positive impacts from a visual metaphor standpoint. Additionally, we planned to gather participants' feedback on their interactions with the prototype as a whole and each view, using a set of six well-designed questions to assess their personal experience. Figure \ref{fig:userstudy} lists specific exploration tasks and detailed questionnaires. Besides, we planned to record their feedback in transcripts via interview for qualitative analysis.

\subsection{Procedure}
After setting up the exploration tasks and questionnaires, we rehearsed a tutorial on how to use \textit{+msRNAer} for exploration, repeated the five exploration tasks, and recorded completion times. We found that each exploration task could be completed within four minutes. Although we conducted the design of \textit{+msRNAer}, the randomness of a force-directed layout in Portrait View did not provide significant benefits in terms of saving our completion time. In other words, every participant could finish each exploration task within four minutes after tutorials. Therefore, we decided to allocate approximately a 60-minute face-to-face session for each participant, consisting of a 10-minute tutorial, approximately 20 minutes of exploration, a fixed 10-minute preset questionnaire, and a 20-minute open-ended interview. Participants were instructed on how to apply \textit{+msRNAer} prototype and were required to complete five specific exploration tasks, with completion times recorded. They were also asked to complete the questionnaire based on their subjective experiences during prototype usage. Additionally, their complementary feedback was also recorded and used for subsequent qualitative analysis.

\subsection{Results}
In this subsection, we discussed the quantitative and qualitative results of our user study.

\textbf{Quantitative Results.}
The quantitative results of our user study were reflected in two aspects. The first aspect was task completion time. We observed that the fluctuations in completion time across these five tasks were within reasonable ranges. The median time to complete the first four exploration tasks varied from 36 seconds to 61 seconds, while the last, more complicated task took a median of 181 seconds to complete. This consequence could be easily explained by the fact that the completion time of tasks E2-E4 was slightly reduced as participants became more familiar with the prototype during the first exploration task. Additionally, the first four exploration tasks focused on simple tasks that could be completed within one or two views. In contrast, the last exploration task required participants to engage with all collaborative views and gradually decipher three interactive results. Consequently, this task demanded more time and attention, leading to a higher median completion time. These results were in line with our expectations, as all tasks were completed within four minutes.

According to the results of the questionnaire, the majority of participants expressed satisfaction with the \textit{+msRNAer} prototype. Specifically, 12 out of 16 participants strongly agreed with its performance, while the remaining four were satisfied with it overall. All participants were highly impressed with the prototype's visual design in Portrait View and expressed varying degrees of satisfaction with its analysis functions, including trends, rankings, and risk factors. Two later questions further highlighted the exceptional integration of both GIS View and MDC View in the prototype. The final question evaluated the implementation of interaction logic, with one participant responding neutrally and the other 15 participants expressing varying degrees of agreement that the interaction implementation was user-friendly.

The quantitative results from two aspects indicated that participants rated the feasibility and effectiveness of the applied exploration tasks and prototype designs relatively highly. They also found the visualization and interaction designs to be intuitive and impressive.

\textbf{Qualitative Results.}
As an extension of the quantitative results, we further employed qualitative analysis during the last 20-minute open-ended discussion. In particular, we conducted opinion research for individuals who did not provide the most satisfactory options on the questionnaire. To derive the qualitative results, we repeatedly reviewed the interview recording, summarized their critical thinking, and achieved a consensus based on their complementary feedback during the interview.

Their feedback was mostly positive with high marks, with only a few critical comments. These comments can be mainly divided into two aspects:

(1) A few participants felt that the current force-directed layout of Portrait View was a "double-edged sword". On the one hand, it provided dynamicity to avoid overlapping and allowed for easy dragging and relocating, but on the other hand, its randomness may increase the workload of recognizing specific portraits in multiple explorations.

(2) Some participants who marked Neutral or Somewhat Agree in Q6 expected to see more interaction logic among views, such as enabling the inclusion of more risk factors in Portrait View or providing solutions for swapping key risk factors with other factors in MDC for higher levels of comparison purposes.

We felt grateful for their feedback as it served as a pre-evaluation before case studies to domain experts and helped us improve this applied prototype. Regarding the comments on the applied force-directed layout, we believe that its advantages outweigh the disadvantages, but we still consider it necessary to improve. As a result, we implemented the search function in GIS View to filter any LGAs or postal areas, which also strengthened the interaction between GIS View and Portrait View to a certain extent.

Since we are proposing a visual modeling method for epidemiological analysis, any other related applications using \textit{+msRNAer} in prototypes may vary in detail. In this situation, with the prior requirements that we consulted with domain experts, we deemed that locking the highest risk factors in each portrait is acceptable because aged, lone, lower-income groups have been proven by domain experts to be high-risk factors, while other objective factors selected from the census were considered to be indirectly related categories. Thus, we explained the reasoning to the participants and derived their understanding and acceptance.

Overall, our user study showed that the design of \textit{+msRNAer} was creative, and the application of \textit{+msRNAer} with the COVID-19 aggregated dataset was also proven to be feasible and effective, although there were some imperfections in certain details. According to the user study, we not only improved \textit{+msRNAer} with search functions in GIS View but also inspired some interesting cases that were demonstrated in future subject-driven case studies.

\section{Case Studies}
\label{sec:case}
Coastal areas are considered densely populated areas prone to cluster infections for transmission, in contrast to the vast land and sparsely populated areas of NSW, Australia. For other incidents, intermittent spreading cycles of virus variants, corresponding policies or restrictions, densely populated residential areas, areas where older people gather, more impoverished areas, or areas with relatively backward public infrastructure may affect the infection situation among different LGAs. In this section, we finalized our \textit{+msRNAer} prototype in two different data aspects based on the COVID-19 aggregated dataset in NSW, which consists of a weekly case summary of each postal area within an LGA in one year and a bi-weekly case summary of LGAs in two years, to the aggregated dataset to provide three prominent cases: Overview-driven cases, Event-driven cases, and Portrait-driven cases, which are based on highlighted driving aspects to compare and explore the significant connections between COVID-19 issues and detailed factors in each LGA census, even attempting to discover the relationships and potential patterns that were really affecting the COVID-19 pandemic behind each LGA portrait. We also combined our findings with facts and news for verification and analysis.

\begin{figure}[h!t]
\centering
\includegraphics[width=3.3in]{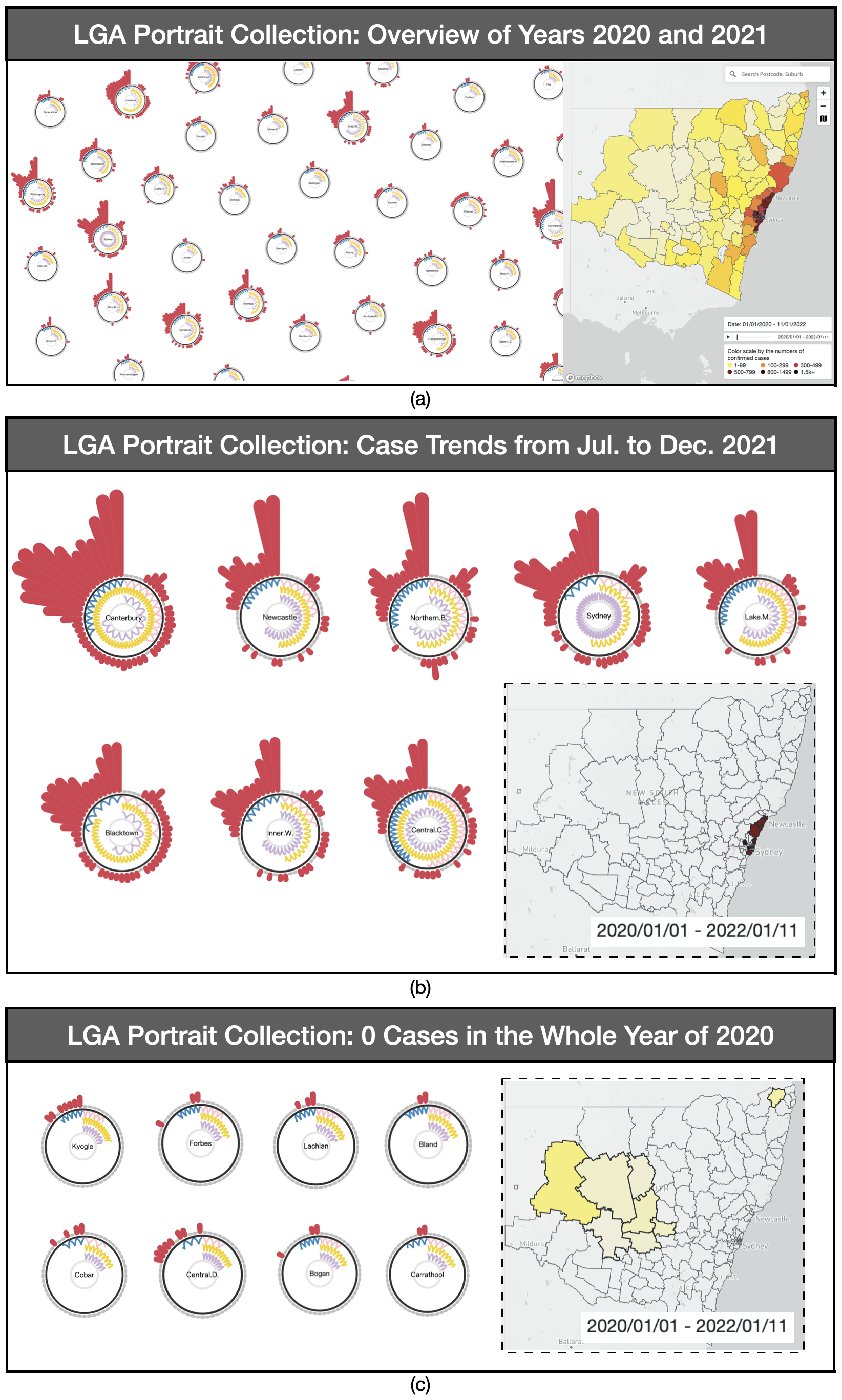} 
\caption{Three cohorts of LGA Portrait Collections: (a) A partial overview of LGA Portraits and geo-locations for 2020 and 2021. (b) Selected LGA Portrait with cases appearing continuously from July to December 2021 (c) Selected LGA Portraits with no cases for the full year of 2020. }
\label{fig:overview1}
\end{figure}

\subsection{Overview-driven Cases}
We first set up a visual representation for the LGA portraits within the period from January 1st to January 11th, 2022. Connected to the GIS View, it showed the severity distribution of COVID-19 cases in geo-polygons. With a quick glance in Figure \ref{fig:overview1}(a), we could distinguish whether the situations are severe or not with roughly two types of portrait appearances. The results are also correspondingly reflected on the map, showing the contrasting perspectives between coastal and inland areas in case numbers.

When comparing one LGA portrait to another, one can examine variations in the COVID-19 cases' spreading situations among LGAs. The S Proteins of each LGA portrait showed the changing trends of COVID-19 cases by height. The only two LGA portrait types that can be distinguished based on appearance are those with COVID-19 cases in both 2020 and 2021 and others with cases only in 2021 but none in 2020. 

We randomly sorted out a few portraits of each type for analysis, as shown in Figure \ref{fig:overview1}(b). The second visual result depicted the trends of COVID-19 cases within an 8-LGA portrait collection. We gained insights demonstrating that they all suffered relatively worse situations and caused two typical waves of cases from mid-year to the end of 2021. Compared to the height of S Proteins in each portrait, they reached the peak of the first wave in September, dropped back to low levels in November, and rose sharply to the highest point in two years. We corroborated the visual result with the actual situation as the NSW government released Delta concerns on July 30th, 2021\cite{NSWDelta}, and the information released on November 28th, 2021, of the first Omicron variant case in NSW\cite{NSWOmicron}. From GIS View, we found they are all located in coastal areas surrounding Sydney City.

The following visual output in Figure \ref{fig:overview1}(c) showed all LGAs with portraits had COVID-19 cases during 2021, although they did not have any cases in 2020. We could further observe that there were 7 LGAs located in inland areas despite the presence of Kyogle LGA near the coastal regions of these 8 LGAs. Compared with cases appearing frequently in other developed LGAs, this consequence was most likely related to the sparse population of these LGAs.

\subsection{Events-driven Cases}
We observed that the first wave of COVID-19 spread across NSW from approximately January to October 2020. Based on the events on the timeline, we divided this period into three phases: uncontrolled, eased, and restricted control. Therefore, we further filtered each LGA portrait into weekly time spans from January 1st to October 20th, 2020. This enabled us to inspect the COVID-19 spreading situations among LGAs during this period, as tracked with the details of these events.

\textbf{Uncontrolled Phase.} After interacting on Control Panel, we hovered the mouse over each E Protein to spot LGAs, which detected the first case occurring in week 4 in Randwick, Paramatta, Kur-ring-gai, and Burwood. 

\begin{figure*}[!t]
\centering
\includegraphics[width=6.85in]{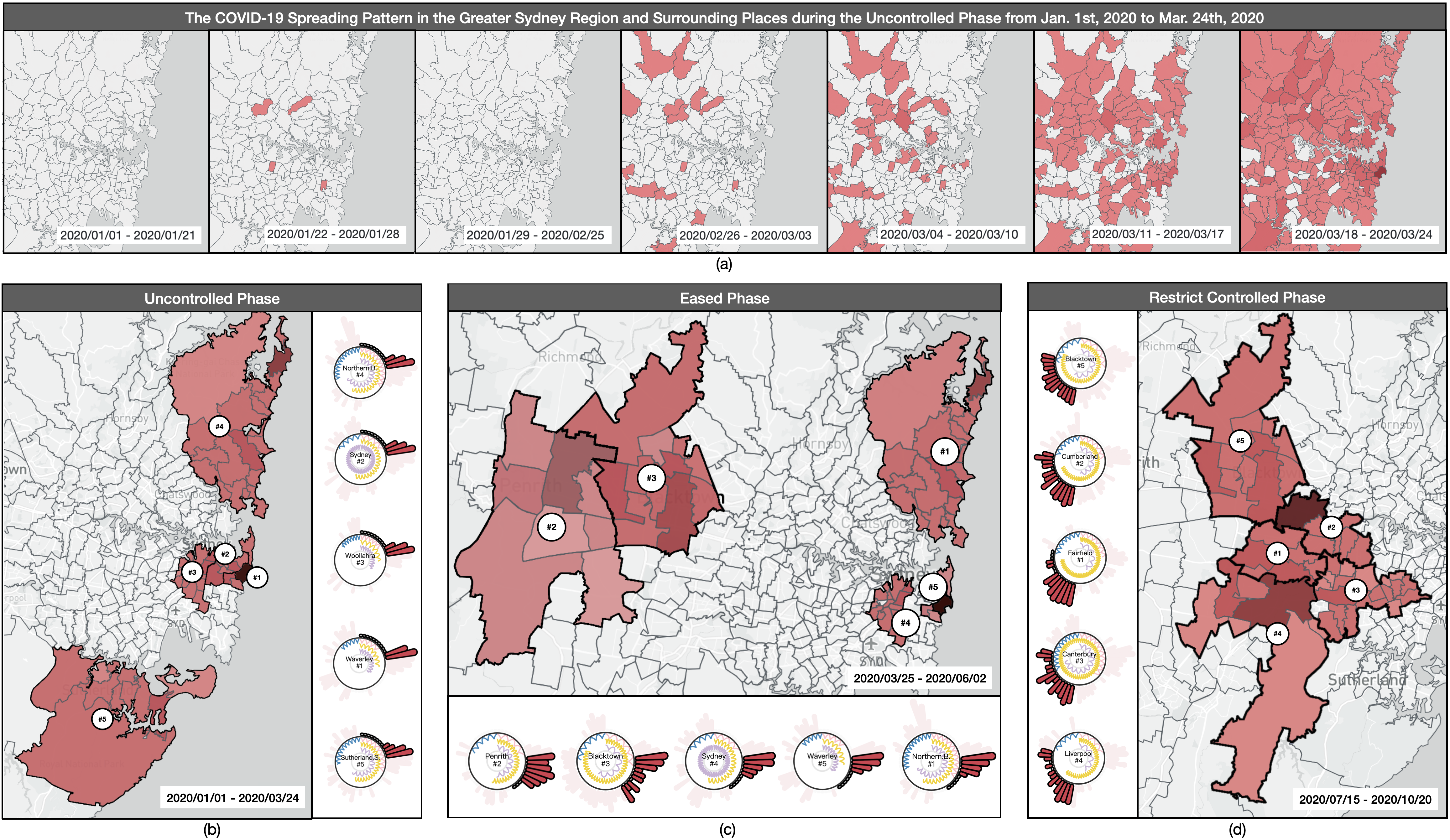} 
\caption{The event-driven cases encompass explorations of (a) the COVID-19 spreading pattern in the uncontrolled phase and  (b)-(d) three phases of overviews. (a) provides a detailed breakdown by time spans of the COVID-19 spreading pattern in the Greater Sydney Region and surrounding places during the uncontrolled phase from 2020/01/01 to 2020/03/24. In (b)-(d), each marked ranking number is arranged according to geo-location, where (b) contains the top 5 LGA portraits in total infection cases with non-controlled events applied to the same phase as (a); (c) displays the top 5 LGA portraits in total in the eased phase from 2020/03/25 to 2020/06/02; and (d) includes the top 5 LGA portraits in total case numbers during the restricted controlled phase from 2020/07/15 to 2020/10/20.}
\label{fig:caseevent1}
\end{figure*}

We hovered the mouse over each E Protein in Control Panel to spot LGA portraits, which could observe the first COVID-19 case occurring in week 4 of 2020 in Randwick, Parramatta, Ku-ring-gai, and Burwood. Concurrently, we noticed the first intervention event started in week 13. Thus, we began by selecting the phase from January 1st, 2020 to March 24th, 2020, to identify the uncontrolled phase. We further utilized the playable timeline window in GIS View to explore the spreading pattern in the Greater Sydney Region and surrounding places. As shown in Figure ~\ref{fig:caseevent1}(a), besides the case that occurred in week 4, there were no other cases during the period from January 1st, 2020 to February 25th, 2020. But starting from week 9 to the end of week 12, cases began spreading in the surrounding places of the Greater Sydney Region, and in a short period, the cases spread severely and finally caused outbreaks in almost all postal areas, especially in the Waverley LGA in darker polygon color.

Drawing an overview of this uncontrolled phase, the visualization results in Figure ~\ref{fig:caseevent1}(b) showed that all LGAs were affected by the first wave of weekly increasing COVID-19 cases, with no restricted events yet. We focused on LGAs with the highest number of COVID-19 infections and interacted with them in GIS View. We located them around the main cities, including Sydney. During this period, Bondi Beach, in the Waverley LGA, had more cases than other LGAs, peaking in only two weeks. We noted that Bondi Beach was crowded with massive gatherings of people, as reported in the news, and the health officials announced a crowd ban on March 21st, 2020. NSW residents were facing physical and psychological pressures due to the spread of COVID-19 and bushfires across the state. These factors indirectly contributed to overcrowding in these tourist hotspots. This situation was also reflected in the Northern Beaches LGA, which saw a rapid increase in COVID-19 cases in the first wave of the pandemic. We further examined the other two LGA portraits, Sydney and Woollahra, and concluded that the consequences were likely due to the connection to Waverley in GIS View, resulting from the movement of people in the adjacent LGA. The fifth infection case in the LGA occurred in Sutherland Shire, located in the southern region of Sydney. We could assume that the COVID-19 pandemic had spread to the outer LGAs at this phase.

\textbf{Eased Phase.} We adjusted the timeline for the first NSW lockdown events from March 25th to June 30th, 2020. The visual results of the top 5 LGAs in Figure ~\ref{fig:caseevent1}(c) showed that the Northern Beaches had the highest number of cases, followed by Penrith and Blacktown, with Sydney LGA dropping to fourth place. Waverley finished fifth. With the special ban in place at Bondi Beach, the number of infection cases in Waverley and its adjacent LGA, Sydney, significantly decreased. We also noted that the NSW lockdown event had a positive impact on LGA cases in the Greater Sydney Region, as reflected by 0 infection cases in the Northern Beaches, Sydney, and Waverley for several weeks in the second half of the selected timeline. Penrith, Blacktown, and Blackburn all had a larger number of low-income residents and are located in the Greater Western Sydney Region, which had more than half the cosine arc of the lower-income RNA and an intense frequency. As a result, we anticipated that the pandemic would spread quickly to other LGAs in the Greater Western Sydney Region.

\textbf{Restrict Controlled Phase.} To test our hypothesis, we further filtered the timeline from July 15th to October 20th, 2020, which was when a new lockdown and tighter restrictions were implemented by the NSW government. The visual results, as shown in Figure ~\ref{fig:caseevent1}(d), reveal that all of the LGAs are connected in the Greater Western Sydney Region. We observed that the most strictly controlled events showed stable increases in cases, even in Cumberland and Liverpool, which had continuous growth in cases over the next few weeks. This confirms our conjecture about the pandemic situation, with a large number of infection cases in the Greater Western Sydney Region.

We also noticed potential patterns among LGA portraits, particularly in the lower-income RNA, demonstrating that all five LGAs have significant lower-income population groups and high population density. In other words, during the implementation of the tightened lockdown, most people will work from home or be self-isolated, and the number of infection cases in the CBD and tourist hotspots will markedly decrease. Thus, the factors in the LGA portraits will dominate and reflect the number of infection cases. Lower-income groups or other population factors in the census may be the most crucial factors affecting the spread of the pandemic.

\subsection{Portrait-driven Cases}
The highlighted interaction in RNAs facilitates the comparison of four key factors within each portrait and multiple attached attributes in MDC View. We selected 12 alternative factors in the census indicators to enable a more comprehensive comparison of the COVID-19 situations based on their portrait factors. To conduct these comparisons, we analyzed portrait-driven cases with different factor influences in collaboration among views.

\begin{figure}[h!t]
\centering
\includegraphics[width=3.3in]{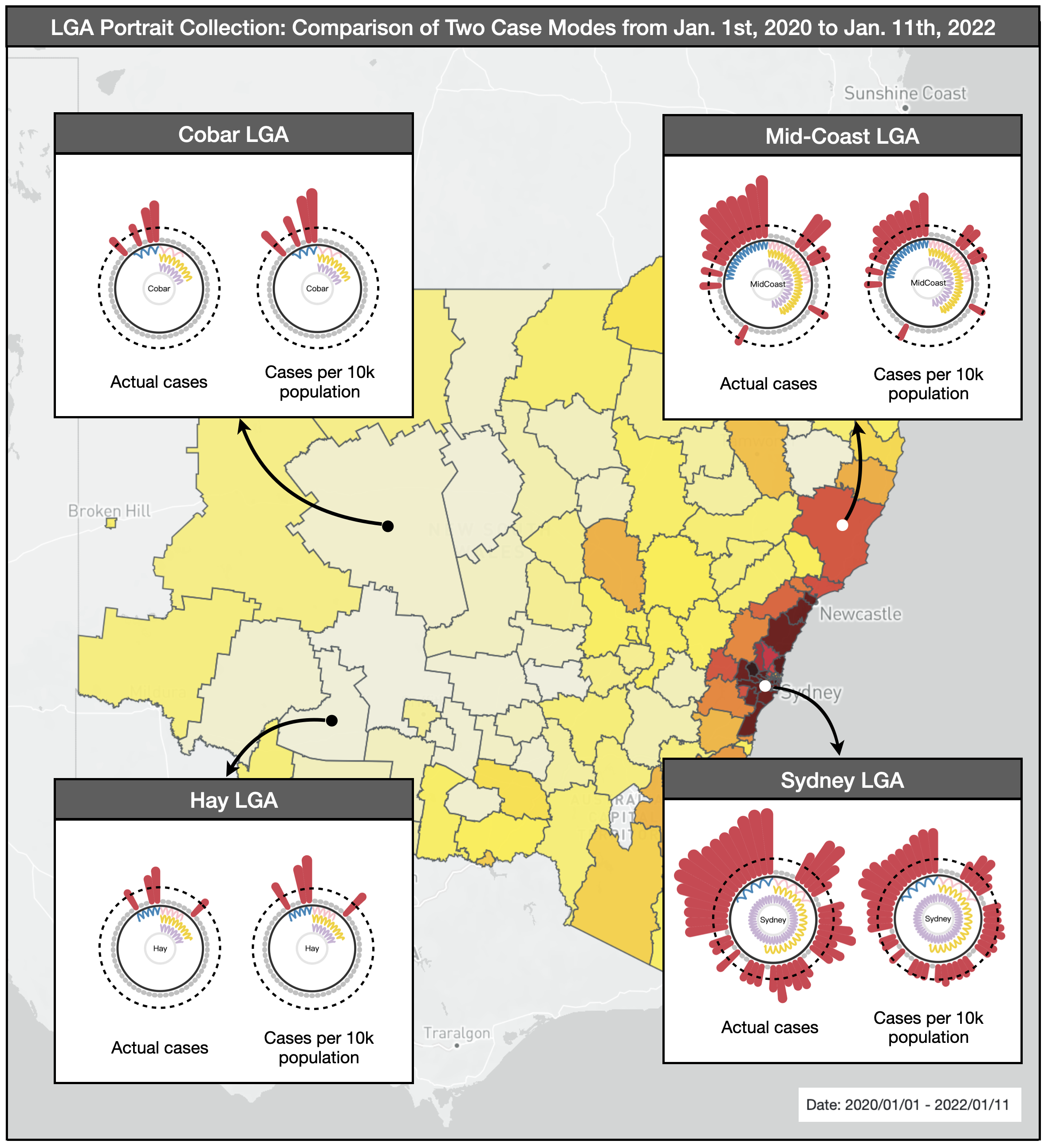} 
\caption{Four cohorts of LGA portraits in different modes (actual cases and cases per 10k population) were selected using LGA polygons in GIS View. Two coastal LGAs (Sydney and Mid-Coast) are marked with white pins, while two inland LGAs (Cobar and Hay) are marked with black pins.}
\label{fig:model1}
\end{figure}

\textbf{Prevalence Rate Influence.}
We began by exploring LGA portraits for the entire time period from January 1st, 2020 to January 11th, 2022 that aimed to compare LGA portraits of coastal areas with inland areas across different case modes. We randomly selected two coastal LGA polygons (Sydney and Mid-coast) and two inland LGA polygons (Cobar and Hay) as representers and compared them using actual cases mode and prevalence rate mode, as depicted in Figure~\ref{fig:model1}.

The consequences conveyed that all S Proteins in Sydney and Mid-Coast LGAs decreased in prevalence rate mode, while all S Proteins in inland LGAs increased in height. The opposite occurred in actual cases mode. We were able to make comparisons among all LGA portraits in both modes since they had been standardized using the same calculation methods in S Proteins' heights.

These findings suggested that almost all LGAs in inland areas did not have enough 10k population, resulting in the height of S Protein in their portraits growing up. We also concluded that the cases per 10k population with standardization were similar in both coastal and inland areas, which validated the strong infectiousness of COVID-19.

\textbf{Case Trends and Lower-income Influences.}
Based on our previous findings, there may be potential relationships between COVID-19 cases and lower-income groups in each LGA. Therefore, we kept the COVID-19 outbreak timeline from the first year (January 1st, 2020 to January 5th, 2021) in the Control Panel and selected the top 10 LGAs with the worst COVID-19 pandemic situations (i.e., the highest total case numbers).

\begin{figure*}[!t]
\centering
\includegraphics[width=6.8in]{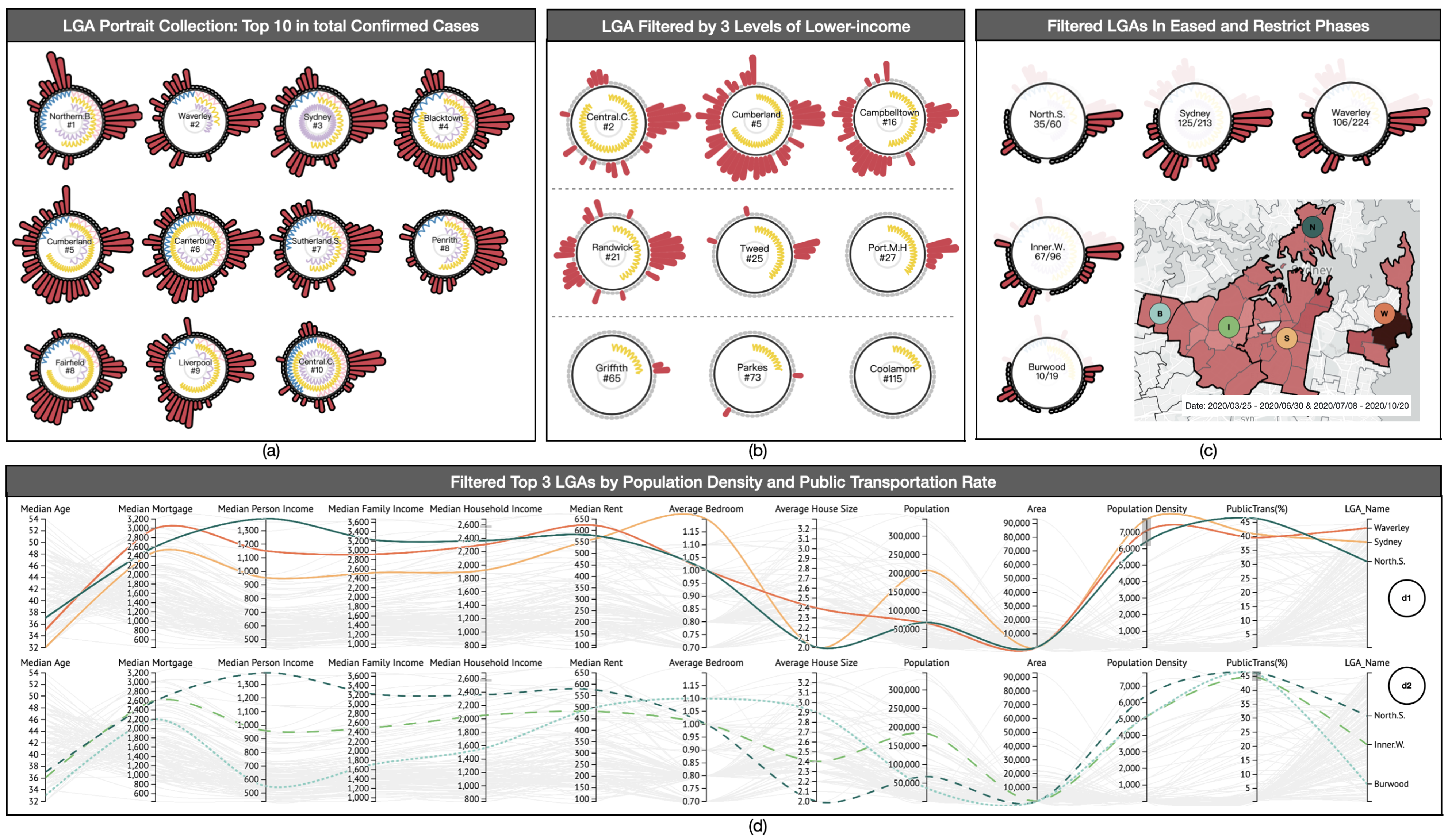} 
\caption{The compiled cohorts of cases are organized according to LGA portrait factors. Specifically, (a) an LGA portrait collection showcases the top ten areas with the highest infection case counts across the entire timeline; (b) a visualization shelf of LGAs is presented in three levels, filtered by lower-income factors; (c) detailed portraits of filtered LGAs are provided, and the colors of map pin-points correspond to the colors used in (d); (d) two portrait cohorts are presented after being filtered by population density and public transportation rate, displayed in multidimensional coordinates, and distinguished by colored lines to differentiate LGAs.}
\label{fig:caseportrait1}
\end{figure*}

From the top 10 rankings displayed on each portrait in the first year, as shown in Figure~\ref{fig:caseportrait1}(a), we identified that areas with RNAs longer than half the range of lengths, or LGAs with the busiest frequency of RNA in one factor, have comparatively worse COVID-19 situations. For example, Fairfield ranked 8th highest in the number of cases with 143 total cases. This area contains the 4th longest lower-income RNA length but the highest tightness frequency. Additionally, the RNA lengths of aged and lone persons in this area do not approach half the length of the full path.

Given that all lower-income RNAs are highlighted in the visual overview, we assume that the lower-income RNA in the LGA portrait is the most influential factor in COVID-19. We selected 9 LGA portraits only based on the appearances of low-income RNA and categorized them into three layers whose lengths are longer than one-half, longer than one-quarter but shorter than one-half, and shorter than one-quarter. As shown in Figure~\ref{fig:caseportrait1}(b), it reveals significant patterns that the COVID-19 situation is getting better by layers, where we scan the length of lower-income RNAs, and they are getting shorter from left to right, top to bottom directions.

\textbf{Combined Influences on LGA portraits, geo-locations, and events.}
We reset the time period to encompass both the eased and restricted controlled phases. Upon analyzing the Greater Sydney Region using GIS View, we observed that adjacent LGAs exhibited similar heatmap polygons. Using an iterative approach, we selected the top 5 LGA portraits with the highest case sums during this period and evaluated their case sums, as well as the total case sums throughout the year. These five LGAs are highlighted in Figure~\ref{fig:caseportrait1}(c), which pin-points correspond to the colors used in Figure~\ref{fig:caseportrait1}(d). With the exception of Waverley LGA, which had been analyzed in previous cases, the other four LGAs were adjacent. While all five LGAs had case ratios exceeding 50\% during the selected period, given the duration of our chosen timeframe, which spanned more than half a year, these eased and restricted events effectively suppressed the spread of COVID-19 in this densely populated region to a certain extent.

\textbf{Household and Dwelling Influence.}
The LGA portraits were reverse-selected to MDC View help to discover the consequences of COVID-19 spread in conjunction with other highlighted census attributes from residents' perspectives. In each column, the median age of each LGA is roughly inversely proportional to financial factors, including the median mortgage, personal income, family income, household income, and rent cost. In most living environments with more than two bedrooms in one household, the majority of people do not meet the standard of one bedroom per person. This indicates that most people still live with others or families, which must also be considered in the analysis of the consequences of the pandemic's spread.

\textbf{Population Density Influence.}
The three LGAs with the highest population densities in the Greater Sydney Region are Sydney, Waverley, and Northern Sydney. These areas have all been severely affected by COVID-19. Upon analyzing their trends on coordinates, we discovered that they share similarities such as a young median age, similar financial profiles, high living costs, and a propensity for using public transit.

\section{Discussion and Interview with Domain Experts}
\label{sec:Discussion}
The \textit{+msRNAer} prototype applied with our aggregated COVID-19 dataset was deployed and used in the workplaces of three domain experts from the Australian government. Through interaction with domain experts, we observed the following:
\subsection {Influence of Key Risk Factors}
\textbf{The combined influence of key risk factors} has a notable negative relationship with the community's resilience profile against the virus spread. According to infection cases, LGAs with a higher representation in four risk groups are all at the top of the list. For example, among the entire list of factors in NSW, the Central Coast has the highest-ranked key factors, ranking \#1 in the aged group, \#1 in the lone person, and \#2 in the lower-income group; Canterbury also has the highest-ranked key factors, ranking \#2 in the aged group, \#1 in the lower-income group, and \#3 in the lone person.

\textbf{The larger the population in the high-risk group, the higher COVID-19 cases}. 

\textbf{Key factor – aged group}: The areas with smaller aged male or female groups among LGAs in NSW have significantly reduced infection cases. In the ranking according to infection cases, the top LGAs generally have older age groups in community representation. There is not much difference in the effect caused by aged male or female groups.

\textbf{Key factor – lower-income}: Among four key risk factors, the lower-income group has the highest impact on COVID-19 risks. The visualization results show that infection cases dramatically increase for LGAs in the lower-income group, roughly occupying one-quarter of the maximum of the lower-income group population in NSW. For example, Cumberland, Canterbury, Sutherland, Sydney, Fairfield, Penrith, Blacktown, Liverpool, and Central Coast are the LGAs with the worst COVID-19 situation, coming with larger population sizes in the lower-income group from small to large.  

\textbf{Key factor – lone-person}: The lone-person group has a certain impact as well. For example, Sydney CBD ranked 3rd according to infection cases and has key risk factors ranking, respectively, lone person (\#1), low-income (\#12), and age group (Male \#25, Female \#29). Compared with other factors across all LGAs, its influence seems the weakest. However, it would also affect LGA's resilience if taking into account other risk factors.

\textbf {Lower risk factors, the higher resilience}: According to the domain expert, if LGAs have a lower presentation of most risk factors, their resilience appears to be higher. This also showed in the situation where several higher resilient LGAs would respond quickly and positively to flatten the curve and reduce the risk, even when they were the areas where these early and severe infections happened during outbreaks.

Domain experts pointed out that these implications largely support the guidelines developed by the Australian government during the pandemic. However, they would also include representation of other under-resourced people in future work, who may also be more at risk of exposure, including aboriginal and Torres Strait Islander people, people living in aged care facilities, and people with disabilities. They plan to share with us more datasets to increase the diversity of higher-risk groups. For the lone-person risk group, domain experts explained that an individual or group's social relationships should be significantly affected and explained that a lone person would have much fewer resources and social support compared to a person living together. Australian and most Western societies encourage young adults to move out and live alone. However, this situation may not be beneficial during the pandemic. Apart from living conditions, they would also like to explore diverse relationship statuses as they believe that residents' relationships might have a greater impact on social support and mental well-being.

\subsection {Influence of Other Factors}
\textbf{The higher population does attract more risk for infection.} All LGAs with larger resident populations should be given more attention during all phases of the pandemic. For example, Canterbury, Blacktown, and Central Coast ranked top 3 in population, and have higher infection cases than other outbreak LGAs.

\textbf {Travel restrictions could effectively control the risk caused by activities related to public transportation}. For instance, North Sydney, Burwood, and Inner West are the top 3 in terms of the number of people who frequently use public transit. These LGAs have the earliest cases during outbreaks before the controlled phase due to the high volume of public transit. However, they respond positively and quickly when the government increases travel restrictions.

\textbf{Geographic factor}: From the overview, the total number of infection cases in coastal areas is generally higher than inland areas for the first 12 months. In particular, those areas with famous beaches had severe infections, for example, the Northern Beaches and Waverley (with Bondi Beach). Analyzing the spreading pattern in Greater Sydney Region during the uncontrolled phase, domain experts observed the outbreak initially appeared in scattered locations throughout the LGA and subsequently spread to areas adjacent to existing cases. During the early stage of the outbreak without control,  spreading patterns were more apparent in residential areas. However, in the latter half of the phase, the pattern became more prevalent in tourist attractions, ultimately resulting in the peak of infection cases in Waverley.

Domain experts have also discovered that the combination of key risk factors and other factors can have a more significant impact. Taking the Northern Beaches as an example, they had the most infection cases in the first year of COVID-19 outbreak, with a greater effect on geographic factors and the highest presentation in lone-person groups, despite a reasonable economic indicator and a low presentation in the elderly group. Although most of these implications seem obvious, they would consider strengthening governance based on these factors, as these implications also prove the effectiveness of government intervention. The experts also emphasized the crucial need for clearer regulations about beaches and recreational usage.

After experimenting with experts, we also conducted a follow-up interview. Overall, experts were impressed by the "actionable insights" that \textit{+msRNAer} obtained. They felt excited about the interface design inspired by the viral 2D structure. They expressed that it supported their memory of each data object and their representation because of the familiarity of the metaphor adopted. They commented, "This would help us reduce the effort and time involved in training other staff." They found that "interaction with humans and interface-in-the-loop is intuitive and assists them in manipulating exploration along a timeline and geographic map." They also used the prototype system to make several valid assumptions and discover unexpected implications. For assumption validation, they found the prototype's capability to help them target higher-risk groups is "specifically useful," and the interactive portrait was "the most useful treasure" considering the magnitude of the data they have.

Experts consistently agreed that the visual modeling of \textit{+msRNAer} can be easily applied in most epidemiological analyses due to the similarity of virus transmission that is often caused by spatiotemporal and objective factors. The prototype will be reusable because the integrated portrait designs helped them augment awareness by porting most data features. They were swiftly able to identify higher risk regions with "constrained factors and higher presentation in risk groups using the realistic". The experts mentioned that they expect the prototype to enable them to understand how to build profiles and predictions for communities in the future based on "reasonable and meaningful information provided."

\section{Limitations and Future Work}
\label{sec:limitation}

\subsection{Limitations}
Our work was constrained by the quality and availability of the data provided, including the related datasets and the redefined risk factors with domain experts. These data issues could have led to inaccuracies in our final results. For example, COVID-19 case data collected by the NSW government is based on usual residence, but some records did not track location accurately, which could have compromised the precision of our analysis. Concurrently, the case data we had access to was recorded by LGAs or postal areas, which limited our ability to examine more granular infection trends in communities. Additionally, our use of a weekly time frame, as dictated by the aggregated datasets, may have limited our ability to uncover more nuanced patterns in the pandemic's progression. The Australian Bureau of Statistics conducted the latest census surveys on communities in 2020, but the complete summary will not be released until mid-2023. Thus, we have utilized the 2016 census in aggregated datasets, which may pose bias in accurately assessing the impact of the pandemic on population demographics and social indicators during this ongoing period. Moreover, the LGA profiles we used contained a vast amount of data, with over 15,000 variables each. We could not concentrate on all variables, necessitating a reduction to only thirteen based on expert supervision. We considered social and economic indicators and indicators of vulnerability to COVID-19, such as rental and mortgage affordability. However, variables in other categories may have directly or indirectly influenced our results. Furthermore, the intervention events in the COVID-19 dataset were constrained by the Australian government system, with different levels of government responsible for administering public policies and programs, which may have resulted in varying intervention measures implemented across LGAs in NSW.

\subsection{Future work}
Based on the feedback from participants in the user study and discussions of case studies with domain experts, we have targeted four aspects to focus on in our future work.

Firstly, we plan to update our prototype by importing refined COVID-19 case data and new census data later next year, when community datasets are collected during the pandemic. 

Secondly, to increase the diversity of community profiles, we will include more variables such as relationships, ethnic groups mentioned by domain experts, and education level. We will continue to apply machine learning algorithms, such as Principal Components Analysis (PCA) for dimensionality reduction to add more categorical indicators and time series predictions for infectious cases affected by risk factors. We also expect to propose a rating system for all variables with assistance and evaluation from domain experts, which would be able to quantify the characteristics of communities. By designing an appropriate measurement matrix for indicators, we will be able to create index metrics for future use in other epidemiological analysis scenarios. These metrics will not only assist decision-makers in making pandemic prevention measurements but will also educate the public on personal influence and, eventually, how to work together to tackle this greatest challenge by putting in their own efforts.

Thirdly, with the improved approach, we intend to conduct a systematic review with more government staff. As suggested, we will offer an automated reporting function for storytelling purposes. This function would help disseminate the right messages to the public.

Finally, we will consider applying our prototype to more transmission datasets in epidemiological analyses to validate the scalability of \textit{+msRNAer}.

\section{Conclusion}
\label{sec:conclusion}
With epidemiological analysis as a backdrop, this paper proposes a new visual modeling method called \textit{+msRNAer} for exploring and comparing spatiotemporal and multidimensional features based on requirements in epidemiology. The method employs a metaphor to assemble portraits that can be used for visualizing each community by combining the visual encoding of time-varying case numbers with objective risk factors that may affect transmissions. The method integrates multiple views, including Control Panel, GIS, and MDC Views, to provide wide-scope observations on filtering events at different severity levels, geo-based spreading distributions, and multidimensional risk factors for each community.

To evaluate the feasibility and effectiveness of \textit{+msRNAer}, we deployed and applied a two-year-scale aggregated dataset by integrating COVID-19 cases with geo-information, NLP-extracted events division on timelines, and risk factors from NSW census Data based on expert supervision. After applying \textit{+msRNAer} into this COVID-19 aggregated dataset, we progressively validated the feasibility and effectiveness of \textit{+msRNAer} by conducting one user study for iterative improving the applied prototype and comparing visual portraits from profuse perspectives in three subject-driven case studies. We summarized how geography, phases of intervention events, and objective risk factors affected COVID-19 spreading situations during the pandemic.

In further interviews with domain experts, we identified additional objective factors that may be influential in the Australian context. We examined pre-existing community factors and discovered practical implications for potential patterns of established community characteristics against the vulnerability facing this pandemic. Despite some limitations and future work, feedback from domain experts suggests that the \textit{+msRNAer} can be considered a common visual modeling method for exploring community-based spatiotemporal and multidimensional features. This method can be applied to abundant epidemiological analyses such as investigating case trends and comparisons, geo-distribution and transmission, risk factors explorations and rankings, and other related tasks.



\section{Declarations}

\subsection*{Availability of data and materials}
We promise that all datasets used in this work are open-source and accessible. The source code is available on Github: \textit{https://github.com/YuDong5018/msRNAers}.

\subsection*{Declaration of competing interest}
The authors have no competing interests to declare that are relevant to the content of this article.

\subsection*{Funding}
This work is supported by National Natural Science Foundation of China (NSFC) under Grant No.61972010 and UTS-CSC Scholarship by the University of Technology Sydney and China Scholarship Council under Agreement No.201908200009.

\subsection*{Authors' contributions}
All authors participated in conceiving, discussing, designing, and writing this work. Yu Dong and Jie Hua made contributions to collecting and processing data. Yu Dong and Christy Jie Liang performed the programming for this visual modeling. Yu Dong, Christy Jie Liang, and Yi Chen put effort into proofreading. Christy Jie Liang and Yi Chen behaved to the corresponding author's duty.

\subsection*{Acknowledgements}
The authors would like to thank all domain experts from the Australian government who provided valuable feedback and comments based on their expertise.

\bibliographystyle{CVMbib}
\bibliography{main}

\subsection*{Author biography}

\begin{biography}[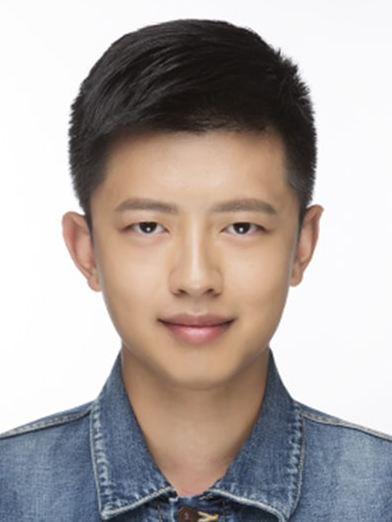]{Yu Dong} received his M.S. degree from Beijing Technology and Business University in 2017. He is currently a Ph.D. candidate at the University of Technology, Sydney. His research interests include data visualization and visual analytics, machine learning, and human-computer interaction.
\end{biography}

\begin{biography}[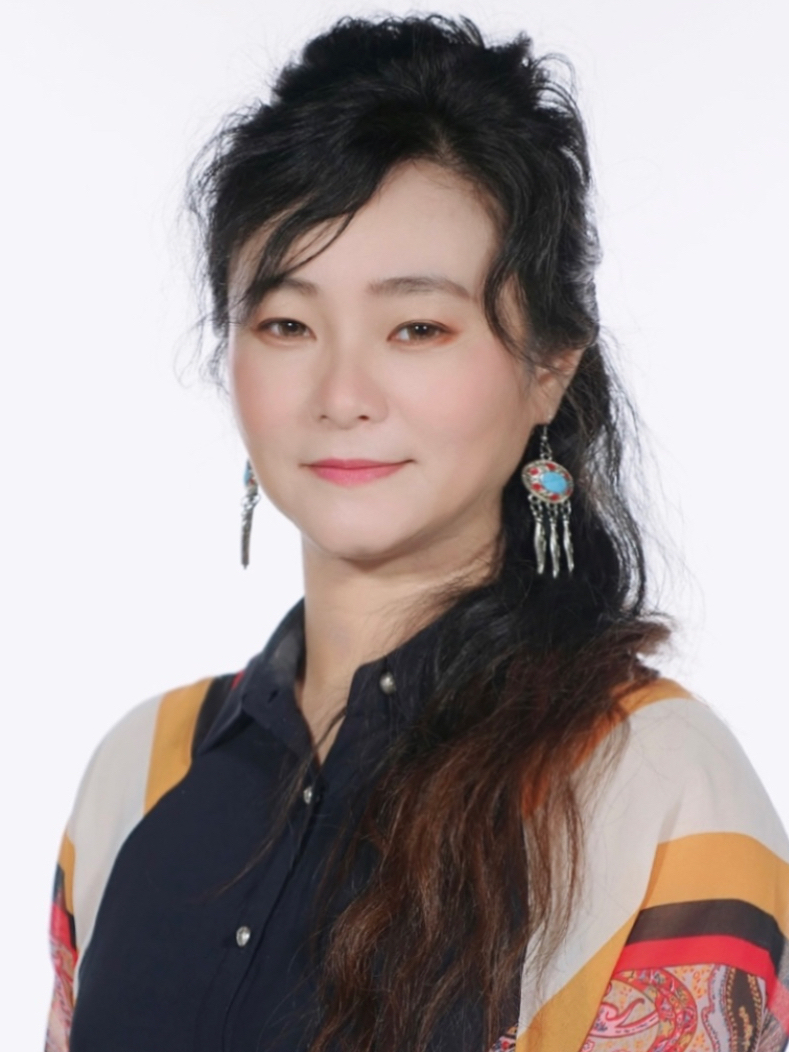]{Christy Jie Liang} received her doctorate in data visualization from the University of Technology, Sydney, and now leads the Data Visualization Research Lab in the Visualization Institute at UTS. Her research interests focus on data visualization and visual analytics.
\end{biography}

\begin{biography}[YiChen.png]{Yi Chen} is a professor and director at Beijing Technology and Business University. She received her Ph.D. degree in Computer Application Technology from Beijing Institute of Technology in 2002. Her research interests include visualization, machine learning, and big data technology for food safety. 
\end{biography}

\begin{biography}[JieHua.png]{Jie Hua} received a Ph.D. degree in software engineering from the University of Technology, Sydney in 2014. He is currently working as a researcher at UTS and Professor at Shaoyang University. His research interests include graph drawing, visualization, big data, and deep learning.
\end{biography}
\vspace*{3.6em}

\subsection*{Graphical abstract}

\begin{figure*}[!t]
\centering
\includegraphics[width=6.8in]{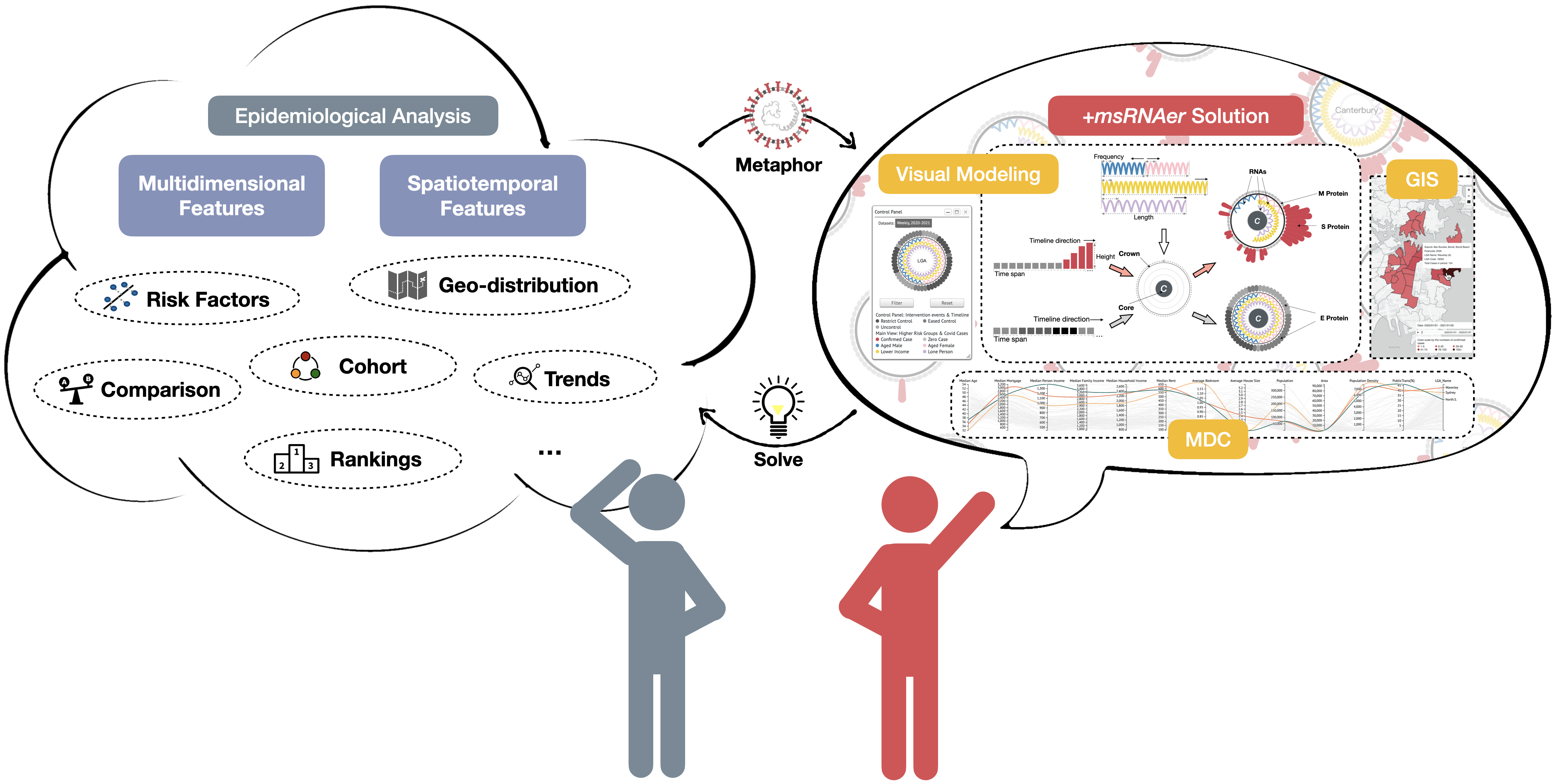} 
\label{fig:graphicalabstract}
\end{figure*}

\end{document}